\documentclass[preprint]{aastex}
\usepackage{epsfig}
\usepackage{natbib}
\usepackage{lscape}
\usepackage{longtable}
\shorttitle{FIR Galaxy Luminosity Function in the Lockman Hole}

\def\f#1   {Figure~\ref{#1}}
\def\s#1   {Sec.~\ref{#1}}
\def\tab#1   {Tab.~\ref{#1}}
\def\t#1   {Tab.~\ref{#1}}
\def\comm#1   {{\tt (COMMENT: #1) }}

\begin{document}

\title{Identification of a Complete 160$\micron$ Flux-Limited Sample of Infrared Galaxies in the ISO Lockman Hole 1-deg$^2$ Deep Fields: 
Source Properties and Evidence for Strong Evolution in the  FIR Luminosity Function for ULIRGs} 
\author{B. A. Jacobs\altaffilmark{1}, 
D. B. Sanders\altaffilmark{1}, 
D. S. N. Rupke\altaffilmark{1,2},
H. Aussel\altaffilmark{3},
D. T. Frayer\altaffilmark{4,5},
O. Ilbert\altaffilmark{1,6},
J. S. Kartaltepe\altaffilmark{1,7}
K. Kawara\altaffilmark{8},
D.-C. Kim\altaffilmark{2,9,10},
E. Le Floc'h\altaffilmark{1,3},
T. Murayama\altaffilmark{11},
V. Smol{\v c}i{\'c}\altaffilmark{12,13,14},
J. A. Surace\altaffilmark{15},
Y. Taniguchi\altaffilmark{16},
S. Veilleux\altaffilmark{2}, and 
M. S. Yun\altaffilmark{17}}
\altaffiltext{1}{Institute for Astronomy, University of Hawaii, 2680 Woodlawn Drive, Honolulu, HI 96822, USA; bjacobs@ifa.hawaii.edu, sanders@ifa.hawaii.edu, drupke@ifa.hawaii.edu}
\altaffiltext{2}{Department of Astronomy, University of Maryland, College Park, MD 20742, USA; veilleux@astro.umd.edu}
\altaffiltext{3}{AIM, Saclay, Bat 709, Orme des merisiers, 91191 Gif-sur-Yvette, France; herve.aussel@cea.fr, emeric.lefloch@cea.fr}
\altaffiltext{4}{NASA Herschel Science Center, California Institute of Technology, MS 100-22, Pasadena, CA 91125, USA} 
\altaffiltext{5}{National Radio Astronomy Observatory, P.O. Box 2, Green Bank, WV 24944, USA ; dfrayer@nrao.edu}
\altaffiltext{6}{Laboratoire d' Astrophysique de Marseille, Marseille, France; olivier.ilbert@oamp.fr}
\altaffiltext{7}{National Optical Astronomy Observatory, 950 N. Cherry Ave., Tucson, AZ, 85719, USA; jeyhan@noao.edu}
\altaffiltext{8}{Department of Astronomy, Graduate School of Science, University of Tokyo, 7-3-1 Hongo, Bunkyo-ku, Tokyo 113-0033, Japan}
\altaffiltext{9}{Department of Astronomy, University of Virginia, P.O. Box 400325, Charlottesville, VA 22904, USA; dk3wc@mail.astro.virginia.edu}
\altaffiltext{10}{National Radio Astronomy Observatory, 520 Edgemont Road, Charlottesville, VA 22903, USA}
\altaffiltext{11}{Astronomical Institute, Graduate School of Science, Tohoku University, Aramaki, Aoba, Sendai 980-8578, Japan}
\altaffiltext{12}{ESO ALMA COFUND Fellow, European Southern Observatory, Karl-Schwarzschild-Strasse 2, 85748 Garching b. Muenchen, Germany}
\altaffiltext{13}{Argelander Institut for Astronomy, Auf dem H\"{u}gel 71, Bonn, 53121, Germany}
\altaffiltext{14}{ California Institute of Technology, MC 249-17, 1200 East California Boulevard, Pasadena, CA 91125 }
\altaffiltext{15}{Spitzer Science Center, California Institute of Technology 220-06, Pasadena, CA 91125, USA; jason@ipac.caltech.edu}
\altaffiltext{16}{Research Center for Space and Cosmic Evolution, Ehime University, 2-5 Bunkyo-cho, Matuyama, 790-8577, Japan; tani@cosmos.phys.sci.ehime-u.ac.jp}
\altaffiltext{17}{Department of Astronomy, University of Massachusetts, 619 Lederle Graduate Research Center, Amherst, MA 01003, USA; myun@astro.umass.edu}

\begin{abstract}
We have identified  a complete, flux-limited, ($S_{\rm 160} > 120$~mJy), sample of 160$\mu$m-selected sources 
from $Spitzer$ observations of the 1-deg$^2$ ISO Deep Field region in the Lockman Hole.  Ground-based UV, optical 
and near-infrared (NIR) 
photometry and optical spectroscopy have been used to determine colors, redshifts and masses for the complete 
sample of 40 galaxies.  Spitzer-IRAC+MIPS photometry, supplemented by ISOPHOT data at 90$\mu$m and 170$\mu$m, 
has been used to calculate accurate total infrared luminosities, $L_{\rm IR}(8-1000\micron)$, and to determine the 
IR luminosity function (LF) of  luminous infrared galaxies (LIRGs).    The maximum observed redshift is $z \sim 0.80$ 
and the maximum total infrared luminosity is log~($L_{\rm IR}/L_{\sun}) = 12.74$.   Over the luminosity range 
log~($L_{\rm IR}/L_{\sun}) = 10-12$,  the LF for LIRGs in the Lockman Hole Deep Field is similar to that found previously for 
local sources at similar infrared luminosities.  The mean host galaxy mass, log~$(M/M_{\sun}) = 10.7$, and dominance 
of HII-region spectral types, is also similar to what has been found for local LIRGs, suggesting that intense starbursts 
likely power the bulk of the infrared luminosity for sources in this range of $L_{\rm IR}$.   However for the most luminous 
sources, log~$(L_{\rm IR}/L_{\sun}) > 12.0$, we find evidence for strong evolution in the LF $\propto (1+z)^{6 \pm 1}$, 
assuming pure number density evolution.   These ultraluminous infrared galaxies (ULIRGs) have a larger mean host 
mass, log~$(M/M_{\sun}) = 11.0$, and exhibit disturbed morphologies consistent with strong-interactions/mergers, and 
they are also more likely to be characterized by starburst-AGN composite or AGN spectral types.  
\end{abstract}


\section{Introduction}
Deep surveys at rest-frame far-infrared (FIR) wavelengths are important for identifying statistically complete samples 
of luminous infrared galaxies (LIRGs: $L_{\rm IR}/L_{\sun} > 11.0$) -- objects that appear to produce the bulk of 
the bolometric infrared luminosity density in the universe, and which are often ``hidden" and misidentified in 
deep UV-optical surveys.   Until recently, progress has been relatively slow in identifying complete samples of 
FIR sources selected at wavelengths $\lambda = 100-200\mu$m, which corresponds to the wavelength 
range where the majority of LIRGs at $z < 1$ emit their peak emission.   The {\it Infrared Astronomical Satellite 
(IRAS)} all-sky survey provided the first complete census of 60$\mu$m-selected galaxies in the local universe 
\citep[e.g.][]{soi89,san03b}, but lacked the sensitivity and long wavelength coverage 
to detect LIRGs at $z>0.05$.   The {\it Infrared Space Observatory (ISO)} provided increased sensitivity and longer 
wavelength coverage, but background instabilities often limited the determination of exact source positions.    
The {\it Spitzer Space Telescope} eventually provided the combination of  long wavelength sensitivity and background stability needed 
to detect sources at the $\sim 100$mJy level with relatively accurate positions, and extensive follow-up observations 
are now underway to identify source properties and redshifts.   

\indent In this paper, we report observations first begun as 
part of the U.S-Japan ISO-ISOPHOT Deep Survey of a $\sim$1-deg$^2$ region 
in the Lockman Hole \citep{kaw98,oya05}, and later 
expanded to include more recent infrared observations of the same 
region obtained as part of the Spitzer Wide-area InfraRed Extragalatic 
(SWIRE) survey \citep{lon03}.   Our final sample consists of MIPS-160$\mu$m  
sources with fluxes greater than 120mJy.  Multi-wavelength follow-up 
observations include Keck spectra of the majority of the sources,  along 
with UV-optical photometry from the Sloan Digital Sky Survey \citep[SDSS:][]{yor00} 
for all of our sources, NIR photometry from the 2-Micron All Sky Survey \citep[2MASS:][]{skr06} 
for most of our sources, and VLA 1.4GHz continuum images of the ISO-ISOPHOT deep fields \citep{yun03}.   
While other surveys \citep[e.g. COSMOS:][]{sco07} may offer superior (particularly ancillary) data, the Lockman Hole 
survey discussed here subtends a total of 1.2~deg$^2$ and is thus comparable in total area to 
similar existing datasets. This survey therefore substantially increases the total area to date at 
these wavelengths and helps guard against cosmic variance.

\indent Sections 2 and 3 describe our field selection and identification of SWIRE 
MIPS-160 sources, respectively.  Methods used for identifying optical counterparts 
are presented in \S~3, along with UV-NIR photometry and optical spectroscopy for individual sources.  
The spectral energy distributions (SEDs) and calculated infrared luminosities ($L_{\rm IR}$) 
for each source are presented in \S~5, and these data are then used to 
determine the Infrared galaxy Luminosity Function (LF).   Section 6 presents evidence 
for evolution in the LF at the highest infrared luminosities, as well as a discussion of the 
properties of the host galaxies, including morphology, colors, masses and spectral types.  
Our conclusions are presented in \S~7.  
\\ \\ 

\section{Field Selection and FIR Observations}
The Lockman Hole (LH) is a region of the sky with the smallest HI column density \citep{loc86} 
and thus has been a favorite target for deep extragalactic surveys, particularly in the FIR  
where confusion noise from infrared cirrus is expected to be at a minimum.   

\begin{figure}[ ] 

\begin{center} 
\includegraphics[scale=0.7]{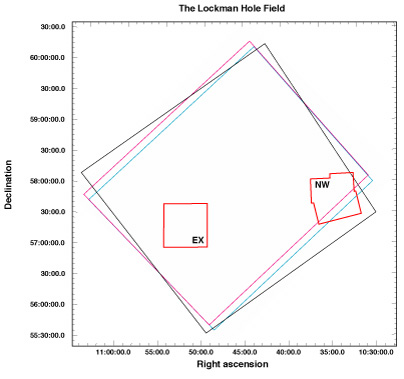}
\caption{ SWIRE and ISO coverage of the Lockman Hole.  The black box denotes the area covered by SWIRE MIPS observations, the magenta box denotes the SWIRE IRAC channels 1 and 3, and the cyan box denotes SWIRE IRAC channels 2 and 4.  The red square marks the location of observations of the ISO LHEX region, the red polygon towards the right of the image marks the ISO LHNW region.}
\label{cov}
\end{center}
\end{figure}

\indent The LH was originally 
targeted for deep FIR observations with the {\it Infrared Space Observatory} (ISO) as part 
of the Japan/University of Hawaii (UH) ISO cosmology program that carried out observations 
of 2 small regions (LHEX and LHNW) using the ISOPHOT detector at 90$\mu$m and 170$\mu$m.  
Both LHEX and LHNW are $\sim44' \times 44'$ in size (see Figure \ref{cov}) - the former is 
centered at $\alpha = 10^{\rm h}52^{\rm m}00^{\rm s}$, $\delta =  57^{\circ}00^{\prime}00^{\prime\prime}$ and 
corresponds approximately to the field center of the {\it ROSAT} X-Ray Deep Survey Field \citep{has98}, 
while the latter is centered at $\alpha = 10^{\rm h}33^{\rm m}55^{\rm s}$, $\delta =  57^{\circ}46^{\prime}20^{\prime\prime}$ 
and was chosen by the Japan/UH team to be centered on the true minimum HI column density 
position within the larger LH field.   Note that a small sky area in the LHNW field was also mapped at 
7$\mu$m using ISOCAM on ISO \citep{tan97} 

\indent There is a fairly extensive published analysis of the ISOPHOT 90$\mu$m and 170$\mu$m 
data for both LHEX and LHNW, including initial source counts \citep{kaw98}, and final 
source lists  \citep{kaw04}, along with attempts to identify radio and near-infrared counterparts for 
individual sources \citep{yun01,san01,yun03}, and to obtain follow-up spectroscopy \citep{san01,oya05}. 
However, several factors prohibited using the ISOPHOT data for producing a well identified 
complete sample of FIR  sources, first of which being the relatively large fluctuations in 
the ISOPHOT background, and the corresponding uncertainty in extracted source positions at 
both 90$\mu$m and 170$\mu$m.   Further complications came with the realization that several 
radio and NIR sources were often found within the relatively large FIR beam. 

\indent  A major improvement in our ability to determine more reliable FIR fluxes and 
more accurate source positions was made possible once data from {\it Spitzer} were made public.    
In particular, SWIRE, a {\it Spitzer} Key Project \citep{lon03} included the LH as one of its 
deep survey areas\footnote{The SWIRE data release paper is available at: \\ http://swire.ipac.caltech.edu/swire/astronomers/publications/SWIRE2\_doc\_083105.pdf}.   SWIRE obtained maps in all 7 Spitzer IRAC+MIPS bands 
(3.6, 4.6, 5.8, 8.0, 24, 70, 160$\mu$m), and covered both LHEX and LHNW, except for 
a small portion of the LHNW region missed by IRAC as shown in Figure \ref{cov}.   

\indent In this paper we use the SWIRE IRAC+MIPS data to produce a new flux-limited sample 
of FIR sources in the LHEX + LHNW fields.  We also make use of all of our previous 
multi-wavelength imaging and spectroscopy along with new data from SDSS and 2MASS 
in order to first identify reliable optical counterparts, and then to determine redshifts and 
total infrared luminosities for each source.  We also use these data to characterize host 
galaxy properties (e.g. masses, morphologies, spectral types) in order to better 
understand the nature of the infrared galaxy population.  
\\ \\

\section{The Lockman Hole 160$\micron$ Sample}
The MIPS-160$\mu$m detector samples emission at wavelengths closest to the rest-frame peak 
of the FIR/submillimeter SED of infrared luminous galaxies.  Most galaxies detected 
at 160$\mu$m are also detected at MIPS-70$\mu$m and MIPS-24$\mu$m as well as all four IRAC 
bands.   For these reasons, as was the case with previous ISO observations of the LHEX 
and LHNW fields, we continue to focus on selecting a complete sample of extragalactic sources at 
the longest observed FIR wavelength. 

\begin{figure}[ ] 

\begin{center} 

\includegraphics[scale=0.8]{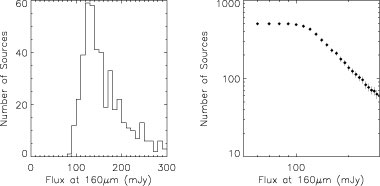}
\caption{Left panel: Distribution of fluxes of the $\sim$500 sources in the SWIRE LH 160$\mu$m catalog.  Each bin is 10mJy wide. Right panel: Integral number counts versus flux of sources in the SWIRE LH 160$\mu$m catalog (horizontal axis ranges from 50 to 300 mJy).  Vertical bars denote Poisson uncertainty.}
\label{complete}

\end{center}
\end{figure}

\indent Figure \ref{complete} (left panel) shows the distribution of flux density of all $\sim$500 
sources in the SWIRE LH 160$\mu$m catalog.  It is from this distribution that we take 120mJy 
as the flux limit for the sample.  There are 39 SWIRE 160$\mu$m sources within the 
LHEX and LHNW fields down to a flux limit of 120mJy.  The integral number counts from the catalog (Figure \ref{complete}, right panel) follow a constant slope at fluxes above the 120 mJy selection limit, and we therefore take the sample to be complete above this limit \citep[e.g.][]{bei88,soi89}.    

\begin{figure*}[] 

\begin{center} 
\begin{tabular}{cc}
\includegraphics[scale=.5]{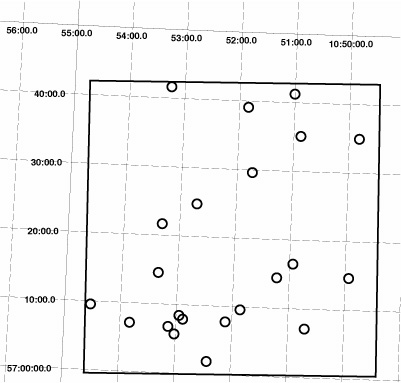}

\hspace{.1in}

\includegraphics[scale=.5]{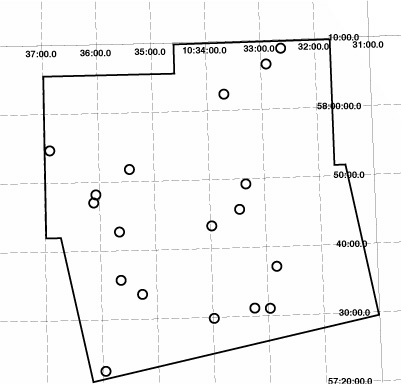}
\end{tabular}
\caption{Left panel: LHEX subregion of the LH.  The coverage of LHEX is roughly: $\alpha = 10^{\rm h}49^{\rm m}30^{\rm s} - 10^{\rm h}54^{\rm m}30^{\rm s}$ and $\delta =  57^{\circ}00^\prime00^{\prime\prime} - 57^{\circ}44^\prime00^{\prime\prime}$.  The circles denote the location of the 160$\mu$m sources. Right panel: LHNW subregion of the LH denoted by the large polygon.  The coverage of LHNW is roughly: $\alpha = 10^{\rm h}31^{\rm m}00^{\rm s} - 10^{\rm h}37^{\rm m}00^{\rm s}$ and $\delta =  57^{\circ}20^\prime00^{\prime\prime} - 58^{\circ}10^\prime00^{\prime\prime}$.}
\label{nwex}
\end{center}
\end{figure*}

\indent  We also compared the MIPS-160$\mu$m catalog with our previous ISOPHOT 170$\mu$m 
catalog, and secondarily with the ISOPHOT 90$\mu$m catalog, and confirmed our previous 
suspicions that a significant fraction of the ISOPHOT 170$\mu$m sources were either spurious 
3$\sigma$ noise peaks or weaker sources with boosted flux due to non-Gaussian noise 
fluctuations.   However, we also discovered a few sources where our previous ISOPHOT data 
along with MIPS-70$\mu$m data suggested that there should have been a MIPS-160$\mu$m 
source above 120mJy, yet there was no source listed in the MIPS-160$\mu$m catalog.  For these cases 
we went directly to the MIPS images and found that in each case there was a source in the 
160$\mu$m image, so we extracted our own flux at 160$\mu$m by performing 
point-spread function (PSF) fitting on the image.  This was done in consultation with the 
SWIRE team so that our flux extraction method would be in agreement with that used to 
construct the SWIRE catalog.  In all but one case our extracted MIPS-160$\mu$m flux 
fell below the 120mJy completeness limit.  However, for J105252.76$+$570753.7, the 
extracted MIPS-160$\mu$m flux of 149mJy was above the completeness limit, thus this source was 
added to the final list, bringing the total number of sources to 40.  The distribution of these 40 sources across the LHNW and LHEX fields is shown in Figure \ref{nwex}.

\indent In addition, there are six sources listed in Table \ref{counter} that were included 
in the SWIRE MIPS-160$\mu$m catalog but not the MIPS-70$\mu$m catalog.  This was somewhat 
surprising given the sensitivity of the MIPS-70$\mu$m catalog, and the expected 70$\mu$m flux 
assuming even a fairly extreme 160/70 flux ratio.  All six sources were in fact visible in the 
70$\mu$m image, and hence, we again performed PSF fitting in consultation with the SWIRE team to
extract a 70$\mu$m flux.   Thus all 40 of our MIPS-160$\mu$m sources now have measured 
MIPS-70$\mu$m fluxes.
\\ \\

\section{Multiwavelength Data}

\subsection{UV/Optical and NIR Images} 
UV/optical and NIR imaging data are critical for identifying the sources   
responsible for the observed FIR emission, measuring redshifts and 
computing luminosities.   The SWIRE data release actually includes optical ($g'r'i'$) 
images obtained at Kitt Peak National Observatory (KPNO) covering a large fraction 
of the LH field,  with extracted photometry for those sources matched to the identified IRAC 
counterparts of MIPS-24$\mu$m sources (see below).  However, SWIRE optical coverage is 
uneven across our fields, a central portion of the LHEX (ROSAT) field was not targeted for 
followup and the southwest portion of the LHNW field falls beyond the edge of the IRAC 
survey (see Figure \ref{cov}) and consequently was also not observed in optical.  As a result, 
almost half (19/40) of our MIPS-160$\mu$m sources lack KPNO $g'r'i'$ photometry
in the SWIRE catalog.   Fortunately, the SDSS provides full 
coverage of the LH field, and we therefore make use of the $ugriz$ catalog photometry from 
Data Release 7 as well as display color-composite images for all of the MIPS-160$\mu$m sources.   
Additionally we use 2MASS catalogs to obtain $JHK_{\rm s}$ photometry for the majority (27/40) 
of our MIPS-160$\mu$m sources. 

\subsection{Identification of Counterparts} 
\begin{figure*}[] \centering
 
\includegraphics[scale=0.5]{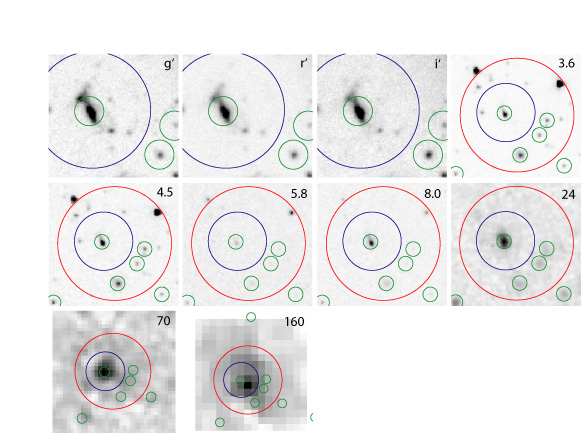}
\caption
{Images of source J103320.32+574913.6 at $g'$, $r'$, $i'$-bands, and 3.6, 4.5, 5.8, 8.0, 24, 70, 160$\mu$m.  The red, blue, and green circles indicate the position of the centroid of each detection in the 160$\mu$m, 70$\mu$m, and bandmerged IRAC+24$\mu$m catalogs, respectively.  The radii of the circles are set to 40\arcsec, 20\arcsec, and 5\arcsec\  in all images (for all sources), and are meant to be rough indicators of the size of the PSF for the 160$\mu$m, 70$\mu$m, and 24$\mu$m sources.  Figures \ref{cut}.1 -- \ref{cut}.40 display cutout images for the entire sample of 160$\mu$m sources and are available at: http://ifa.hawaii.edu/\textasciitilde{}bjacobs/LHonlinefigs.pdf}
\label{cut}

\end{figure*}

As an initial step the 160$\mu$m sources were matched with their
counterparts in the SWIRE 70$\mu$m, and IRAC+24$\mu$m+Optical
catalogs.  A series of image cutouts from each of the Spitzer IRAC+MIPS bands 
plus optical images from KPNO  were then assembled for each source 
in order to identify the correct optical counterpart (see Figure \ref{cut} for an example\footnote{Images of the complete sample are available at: http://ifa.hawaii.edu/\textasciitilde{}bjacobs/LHonlinefigs.pdf}).  The red, blue, and green circles overlaid
on these images indicate the position of the centroid of each
detection in the 160$\mu$m, 70$\mu$m, and bandmerged IRAC+24$\mu$m
catalogs, respectively.  Each 160$\mu$m source has one and only one counterpart in the 70$\mu$m images, and nearly all (36/40) have only one counterpart at 24$\mu$m.  In the cases with multiple 24$\mu$m sources, we choose the source nearest the 160$\mu$m centroid.  Table \ref{counter} lists the source identification numbers from the SWIRE catalogs, as well as coordinates of each source taken from the SWIRE IRAC+24$\mu$m+Optical catalog, except for sources lacking coverage in two or more IRAC bands (noted in the table) where we list coordinates from the SWIRE 24$\mu$m catalog.  Following the identification of the Optical/IRAC counterpart
for each of the MIPS-160$\mu$m sources, we then match these with objects in the 2MASS and SDSS 
catalogs and list their source IDs.  In the case of 2MASS we prefer to quote the 
Extended Source Catalog (XSC) when a match is available, but also make use of 
the Point Source Catalog (PSC).   The last two columns in Table \ref{counter} list 
the IDs assigned to sources by \citet{oya05} using observations from ISO.  Table \ref{fltab} 
lists the fluxes reported in the SWIRE, 2MASS, and SDSS catalogs along with our own 
flux measurements as described above. 

\subsection{Optical spectra}
\begin{figure*}[] 

\begin{center} 

\includegraphics[width=\textwidth]{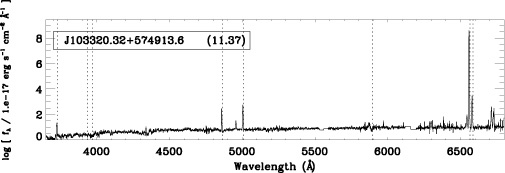}

\caption{Example spectrum of J103320.32+574913.6 (see Table \ref{lumtab} for the source of the data).  The wavelength range is 3650 - 6800\AA, and the vertical lines mark the positions of [OII]3726/3729, CaII H \& K, H$\beta$, [OIII]5007, NaI D, H$\alpha$, and [NII]6583.  Figures 5.1-- 5.28 display spectra for the 28 sources for which we have data, see: http://ifa.hawaii.edu/\textasciitilde{}bjacobs/LHonlinefigs.pdf}
\label{spec}

\end{center}
\end{figure*}

We collected spectroscopic data on the optical counterparts to the ISO
sources from several sources, first of which was our library of Keck spectra 
obtained as part of the original ISO follow-up program which had targeted 
expected counterparts to the ISOPHOT 170$\mu$m sources.    Both low- and
moderate-resolution spectra were taken with the Echellette
Spectrograph and Imager \citep[ESI:][]{she00} over several observing
runs: 2000 Mar $30-31$ and 2001 Jan $23-24$ UT for the low-resolution
data; and 2001 Feb $27-28$, 2002 Jan $16-17$, 2002 Feb 16, and 2002
Mar 15 for the moderate-resolution data.  The low-resolution spectra
had previously been used for redshift identification of putative ISOPHOT 170$\mu$m 
counterparts and some of these data were published in \citet{oya05}.   The follow-up
high-resolution spectra were obtained in order to make accurate emission-line flux
measurements.  These are published here for the first time.  In total, we have Keck spectra for  
19/40 (48\%) of our complete sample.  An example of our high-resolution Keck spectra 
is shown in Figure \ref{spec}.    In addition to Keck, we supplemented our redshift and flux 
measurements with public SDSS spectra \citep{ade07} for an additional 9 of our 
targets.    All 28 of the Keck + SDSS spectra are published on-line in Figures \ref{spec}.1-\ref{spec}.28\footnote{See: http://ifa.hawaii.edu/\textasciitilde{}bjacobs/LHonlinefigs.pdf}. 
In addition we list spectroscopic redshifts for 2 sources  for which the 
spectra themselves are unavailable (see Table \ref{lumtab}).  

\indent Ten of our MIPS-160$\mu$m sources do not have optical spectra.  Most of these 
sources were either not properly identified in the earlier ISOPHOT-170$\mu$m images, or had several 
radio and K-band counterparts within the ISOPHOT-170$\mu$m beam where the dominant 
counterpart that had previously been targeted with Keck turned out not to be the correct source.  
For these 10 sources, we have been able to determine fairly accurate photometric redshifts, as 
described below.  

\subsection{VLA 1.4 GHz Radio Data}

Deep 1.4 GHz radio continuum images of the LHEX and LHNW fields were obtained using the NRAO VLA in the B-configuration in February 2000 and March-April 2001 as part of the AY110 and AY121 programs.  The angular resolution of the data are $\sim5\arcsec$.  The achieved sensitivity of the LHEX data is $1\sigma\sim15$ $\mu$Jy while the sensitivity for the LHNW field is a factor of 2 worse because of a bright (4.2 Jy) continuum source nearby.  The photometry is done using the AIPS task SAD, which fits a 2D Gaussian to the brightness distribution, and total integrated flux is reported for extended sources in Table~\ref{fltab}.  A more detailed discussion of the radio data is presented by \citet{oya05}.

\section{Results}
The complete photometric data set for each of our 40 MIPS-170$\mu$m sources is 
presented in Table \ref{fltab}.   We use these data, along with the measured (spectroscopic) and computed 
(photometric) redshifts listed in Table \ref{lumtab}, to construct SEDs for each source 
and to compute total infrared luminosities, which are then used to construct the luminosity function
for our complete sample.

\subsection{Photometric Redshifts}
The majority of the SWIRE 160$\mu$m detections have spectroscopic data, which were used to determine their redshifts.  
  For the 10 objects without spectra (noted with ``Phot-z'' in Table \ref{lumtab}), we calculate a photometric
redshift using the photometry from SDSS $ugriz$-bands (except for J103341.28+580221.4, see Table \ref{lumtab}), 
2MASS $JHK_{\rm s}$-bands and IRAC 3.6 and 4.5$\mu$m.  
A $\chi^2$ template-fitting method ({\it Le Phare}) was used following the prescription given in \citet{ilb09}.  This method 
offers an improvement in photo-z accuracy over previous methods, due primarily to improved calibration using 
large spectroscopic samples from VLT-VIMOS and Keck-DEIMOS.  The best fit redshift values and uncertainties are 
listed in Table \ref{lumtab}. 

\subsection{Spectral Energy Distributions and Infrared Luminosity}
\begin{figure*}[] 

\begin{center}
\includegraphics[scale=.9]{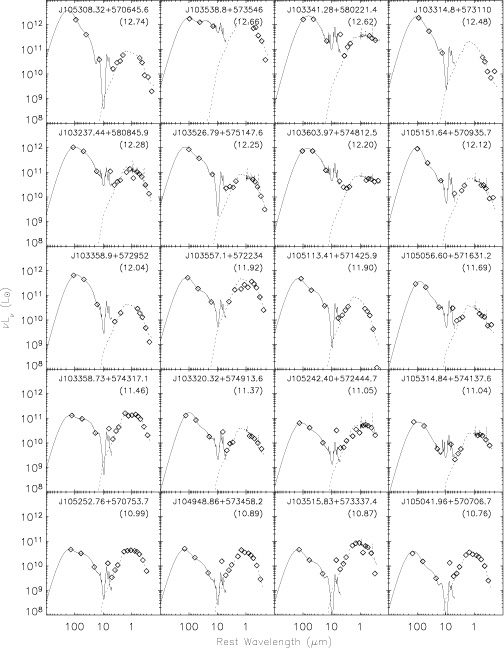}

\caption{\scriptsize SEDs of SWIRE 160$\mu$m sources in $\nu L_{\nu}$ in units of solar luminosity versus rest wavelength (increasing to the left).  The solid line represents a \citet{sie07} model fit to the MIPS and 8$\mu$m data, shown as diamonds.  The dotted line represents the stellar component of the luminosity and is fit to the shorter wavelength data \citep{ilb10}.  Sources are sorted by decreasing IR luminosity.  We list the total-IR luminosity in each panel in units of log$(L_{\rm IR}/L_{\sun})$ in parentheses below the source ID.}
\label{seds1}

\end{center}
\end{figure*}

The photometry and redshifts were used to construct SEDs ($\nu L_{\nu}$) for each source, which are shown 
in Figure \ref{seds1} sorted in order of decreasing luminosity.   The SEDs are characteristic of what has 
previously been observed for infrared-selected galaxies, with the most luminous sources showing a dominant 
``infrared bump" presumably due to thermal dust emission, and an ``optical bump"  due to thermal emission from stars.  Although the 
mid-infrared sampling is relatively sparse, it is also possible to see the effects of emission from polycyclic aromatic hydrocarbons 
(PAHs) in the mid-infrared at $\lambda_{\rm rest} \sim 4-12\mu$m, and silicate absorption at $\lambda_{\rm rest} \sim 10\mu$m. 

\indent The SEDs displayed in Figure \ref{seds1} also show template fits to the data.  The MIPS and 8$\mu$m 
points are fit to a library of SED templates by \citet{sie07}, and the best fit is shown as solid line.  
The dotted line in Figure \ref{seds1} represents a stellar evolution model fit to the UV-NIR data which is used to 
estimate stellar masses (see below).
To estimate each source's total IR luminosity, $L_{\rm IR}(8 - 1000\mu$m), we use the prescription described by \citet{sie07}.  
The use of this model of SED fitting to estimate IR luminosity over others, such as \citet{cha01} or \citet{dal02}, is advocated 
by \citet{sym08} largely due to the tendency of these models to underestimate the peak of the FIR luminosity as 
represented by the 160$\mu$m flux.  This tendency of \citet{sie07} to better fit the FIR data held true for our 160$\mu$m 
sample as well.  Table \ref{lumtab} lists the computed infrared luminosity for each source.

\subsection{Infrared Galaxy Luminosity Function}
\begin{figure} \centering
 
\includegraphics[scale=0.75]{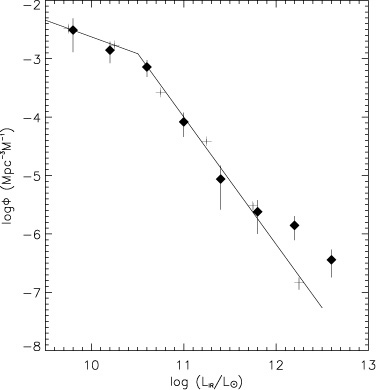}
\caption
{Infrared galactic LF from observations of the LH by SWIRE are shown as filled diamonds with 68\% uncertainty bars. The RBGS data are shown as pluses fit  by two power-laws \citep{san03b}.  At log$(L_{\rm IR}/L_{\sun}) = 9.5 - 10.5$ the RBGS is fit with $\Phi (L) \propto L^{-0.6 \pm 0.1}$, and at log$(L_{\rm IR}/L_{\sun})= 10.5 - 12.5$  the power-law is: $\Phi (L) \propto L^{-2.2 \pm 0.1}$.}
\label{lf}

\end{figure}

In addition to infrared luminosity, we use each source's flux at 160$\mu$m, and redshift 
($H_{0} = 75$ km s$^{-1}$ Mpc$^{-1}$, $\Omega_{\rm m} = 0.3$ and $\Omega_{\Lambda} = 0.7$ to 
calculate luminosity distance) to calculate a LF from our sample.   Figure \ref{lf} shows the infrared LF 
resulting from our observations of the LH in comparison with the LF in the local Universe, 
previously determined from the IRAS Revised Bright Galaxy Sample (RBGS) all-sky survey, which has 
median and maximum redshifts: $z=0.008$ and $z=0.09$ \citep{san03b}.  
The LF density values and uncertainties plotted in Figure \ref{lf} are listed in Table \ref{lfparam}.  
We divide the data into bins of log$(L_{\rm IR}/L_{\sun}) =0.4$ in size, corresponding to steps of one in 
absolute magnitude.  Note that SWIRE detected galaxies, J103258.0+573105 and J105349.60+570708.1 
at 160$\mu$m, but we do not include them in our analysis because their calculated total infrared luminosities 
are log$(L_{\rm IR}/L_\sun) = 9.55$ and 8.81 which results in them falling as the lone galaxy in their respective 
luminosity bins.  The volume at which the survey is sensitive to galaxies below this luminosity range is small, 
so we restrict our attention to higher luminosity sources.

\indent In comparing the LF of our LH data with the RBGS we are comparing a narrow deep survey with 
40 galaxies to a wide local survey with several hundred galaxies.  This means that the relative significance of 
each galaxy is higher for the LH, so it is important to have accurate luminosity estimates.  On the 
other hand, the one-magnitude bins have the effect of mitigating uncertainties in luminosity, since they can 
include galaxies with luminosity estimates differing by as much as a factor of 2.5.  This is particularly significant 
for the objects in the sample that have luminosity estimates using photometric redshifts, since their luminosity 
is less certain than the rest of the sample.   The photometric redshifts and their 68\% uncertainty limits are noted 
in Table \ref{lumtab}, and these limits are used to estimate uncertainties in their luminosities.  In addition to 
photometric redshift uncertainties, the errors in matching $\nu L_{\nu}$ to a model infrared luminosity become 
less important when the data are binned as described.   The space density of galaxies within each bin is calculated 
using the method developed by \citet{sch68}, which accounts for the fact that in a flux-limited sample a larger range 
of luminosities is observable at small distances than at greater distances.  He proposes using a measure of the volume that a 
particular flux measurement samples, given the flux limits of the survey.  For example, a galaxy at redshift 
$z=0.5$ with a flux at 160$\mu$m of 170mJy could have been seen at greater redshift (and hence represent 
a larger volume), since the flux limit of the sample is 120mJy.  This volume sampling is characterized by the 
$V/V_{\rm max}$ parameter, where $V$ is the volume corresponding to the redshift actually observed, and 
$V_{\rm max}$ is the maximum volume over which it could be observed.   A mean $V/V_{\rm max}$ value of 
0.5 within a luminosity bin indicates an even distribution of galaxies within the total volume sampled in that bin.
\\ \\

\section{Discussion}

\subsection{Evidence For Possible Evolution in the Luminosity Function}
The space density of galaxies with infrared luminosity (8 - 1000$\mu$m) in the range log~$(L_{\rm IR}/L_{\sun}) = 10 - 12$ 
appears to be consistent between the RBGS and our LH sample.  In particular, \citet{san03b} fit a broken power-law to the 
RBGS sample.  At log~$(L_{\rm IR}/L_{\sun}) = 9.5 - 10.5$ the RBGS is fit with $\Phi (L) \propto L^{-0.6 \pm 0.1}$, and at 
log~$(L_{\rm IR}/L_{\sun})= 10.5 - 12.5$  the power-law is: $\Phi (L) \propto L^{-2.2 \pm 0.1}$.  The Lockman Hole data 
for luminosities log~$(L_{\rm IR}/L_{\sun}) < 12$, agree within their errors to these power-laws.  This concurrence 
is to be expected given the relatively low redshifts sampled in these lower luminosity bins.  At log~$(L_{\rm IR}/L_{\sun}) > 12.0$ 
the situation changes.  The co-moving space density of ultraluminous infrared galaxies (ULIRGs: log~$(L_{\rm IR}/L_{\sun}) > 12.0$) 
in the log~$(L_{\rm IR}/L_{\sun})= 12.0 - 12.4$ luminosity bin is $\sim 7\times$ higher in the LH than in the RBGS.  
The median redshift of the ULIRGs in the LH in this luminosity bin is $z = 0.51$.  In the highest luminosity bin, 
log~$(L_{\rm IR}/L_{\sun}) = 12.4 - 12.8$, the median redshift of the 4 LH galaxies in this bin is $z = 0.71$. To compare the co-moving space density with the RGBS in this bin we extrapolate the RBGS power-law to log~$(L_{\rm IR}/L_{\sun}) = 12.6$ and find that the density in the LH sample is $\sim 11\times$ higher.

\indent Our new results for the LF of the most luminous infrared 
galaxies in the LH are consistent with strong evolution in the co-moving space density of ULIRGs.    
If we assume pure space-density evolution of the form $(1+z)^{n}$, our new results for the LH imply $n \sim 6 \pm 1$.   
This is similar to what was found in an earlier study of the infrared luminosity function of  ULIRGs by \citet{kim98}, 
where the co-moving space density of ULIRGs in the IRAS 1-Jy sample (mean $z \sim 0.15$), was found to be 
$\sim2\times$ larger than the local space density of ULIRGs in the RBGS (mean $z \sim 0.05$), implying  $n = 7.6 \pm 3.2$ .  
Our new results are also consistent with a recent determination of the extragalactic 250$\mu$m luminosity function 
by \citet{dye10}, which shows a ``smooth increase" with redshift of a factor of 3.6$\times$ in the co-moving space 
density of luminous infrared sources between $z = 0$ and $z = 0.2$, corresponding to $n = 7.1$.  

\indent Deeper far-infrared surveys currently underway with $Spitzer$ and $Herschel$ will eventually allow us to 
determine whether the strong evolution observed for the most luminous infrared extragalactic sources in the relatively nearby universe 
continues out to higher redshift.  For now, we simply note that if we assume similar strong evolution, e.g.  $(1+z)^6$, in the 
ULIRG population out to higher redshifts, our results would imply a co-moving space-density of ULIRGs that is 
$\sim$700$\times$ larger at $z \sim 2$ compared to the value at $z = 0$.   Is there evidence for such a large population of 
ULIRGs at high redshift?  The answer seems to be yes.  There is a population of faint submillimeter sources detected by 
the Submillimeter Common User Bolometer Array (SCUBA) on the James Clerk Maxwell Telescope (JCMT), which has 
been interpreted variously as exotic objects, or ULIRGs at high redshift \citep{sma97, hug98, bar98, lil99}.  \citeauthor{lil99} 
argued that these objects are indeed ULIRGs at $z \sim 2$.  Subsequently, \citet{cha05} measured a range of spectroscopic redshifts, $z=1.7-2.8$ for a sample of 73 submillimeter galaxies, and suggested an evolution in number density of three orders of magnitude for ULIRGs between $z=0$ and $z\sim 2.5$.  Our results for ULIRGs in the LH, when extrapolated out to $z = 2-2.5$ are then consistent with 
the hypothesis that the SCUBA submillimeter sources are indeed ULIRGs.

\subsection{Galaxy Properties}
To achieve a better understanding of the processes responsible for the observed infrared emission and the 
nature of the galaxies in our MIPS-160$\mu$m sample, we use our UV-NIR imaging data and 
optical spectra to determine galaxy morphology and masses, and spectral types, respectively. 

\subsubsection{Imaging: Morphology and Masses}
\begin{figure*}
\begin{center}
\begin{tabular}{cccc}

\scriptsize J105308.32$+$570645.6, (12.74)&\scriptsize J103538.8$+$573546*, (12.66)
&\scriptsize J103341.28$+$580221.4*, (12.62)&\scriptsize J103314.8$+$573110*, (12.48)\\
\includegraphics*[scale=.65]{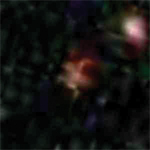} &
\includegraphics*[scale=.65]{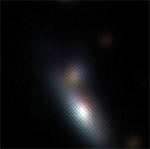} &
\includegraphics*[scale=.65]{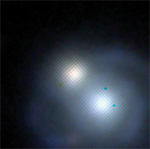} &
\includegraphics*[scale=.65]{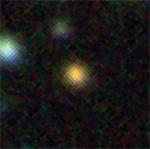} \\
\scriptsize J103237.44$+$580845.9, (12.28)&\scriptsize J103526.79$+$575147.6*, (12.25)
&\scriptsize J103603.97$+$574812.5*, (12.20)&\scriptsize J105151.64$+$570935.7*, (12.12)\\
\includegraphics*[scale=.65]{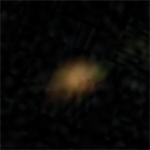} &
\includegraphics*[scale=.65]{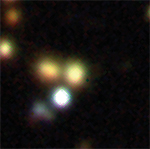} &
\includegraphics*[scale=.65]{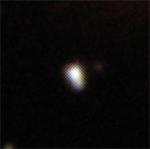} &
\includegraphics*[scale=.65]{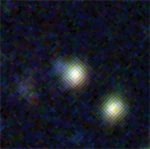} \\
\scriptsize J103358.9$+$572952*, (12.04)&\scriptsize J103557.1$+$572234, (11.92)
&\scriptsize J105113.41$+$571425.9*, (11.90)&\scriptsize J105056.60$+$571631.2*, (11.69)\\
\includegraphics*[scale=.65]{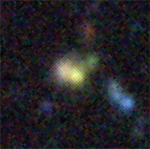} &
\includegraphics*[scale=.65]{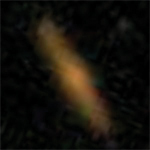} &
\includegraphics*[scale=.65]{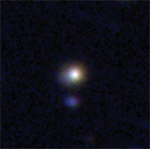} &
\includegraphics*[scale=.65]{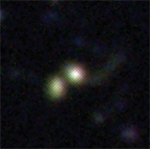} \\
\scriptsize J103358.73$+$574317.1, (11.46)&\scriptsize J103320.32$+$574913.6, (11.37)
&\scriptsize J105242.40$+$572444.7, (11.05)&\scriptsize J105314.84$+$574137.6, (11.04)\\
\includegraphics*[scale=.65]{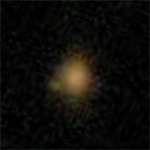} &
\includegraphics*[scale=.65]{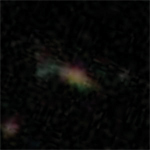} &
\includegraphics*[scale=.65]{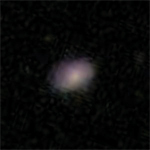} &
\includegraphics*[scale=.65]{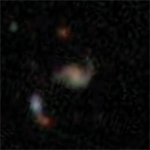} \\
\scriptsize J105252.76$+$570753.7, (10.99)&\scriptsize J104948.86$+$573458.2, (10.89)
&\scriptsize J103515.83$+$573337.4, (10.87)&\scriptsize J105041.96$+$570706.7, (10.76)\\
\includegraphics*[scale=.65]{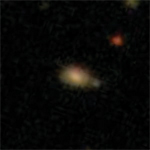} &
\includegraphics*[scale=.65]{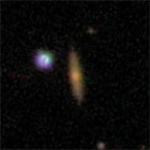} &
\includegraphics*[scale=.65]{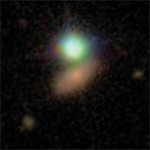} &
\includegraphics*[scale=.65]{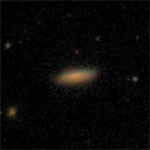} \\

\end{tabular}
\end{center}
\caption{\scriptsize
 Stacked images of the LH sources sorted by luminosity.  Most images are from SDSS finding charts, however for sources at $z > 0.3$ we show stacked KPNO $g'$+$r'$+$i'$ images when available, because these reveal more detail than SDSS (these images are marked with an asterisk).  Images are oriented with North up and East to the left, with dimensions of 100 kpc $\times$ 100 kpc. We list the total-IR luminosity for each source in units of log$(L_{\rm IR}/L_{\sun})$ in parentheses following the source ID.}
\label{grij}
\end{figure*}

In order to develop a picture of the morphologies, and to gain an indication of the prevalence of merging/interacting galaxies 
in the sample we examine their UV-NIR images.  We compile color composite images of the sources from those 
available through the Finding Chart section of the SDSS DR7 website, and show these in luminosity order in Figure \ref{grij}.  
The zoom on these cutouts is scaled so that each box is 100kpc on a side.  At high redshifts and thus high zoom, the image 
quality of the SDSS charts is low, so for sources with $z > 0.3$ we display stacked $g^\prime r^\prime i^\prime$ images from KPNO when available.  
A brief description of the galaxy morphologies is presented in Table \ref{morph}.  Many of the higher luminosity sources with 
log$(L_{\rm IR}/L_{\sun}) > 11.5$ exhibit features suggestive of interactions/mergers, such as multiple cores and/or tidal tails.  
At luminosities lower than log$(L_{\rm IR}/L_{\sun}) < 11$  the large majority of sources appear to be mostly unperturbed spirals.  
These trends are consistent with previous studies of local samples of  LIRGs and ULIRGs \citep[e.g.][]{san96}, which have 
shown that strong interactions and mergers appear responsible for triggering the most luminous infrared sources. 

\indent Stellar masses for each of the MIPS-160$\mu$m sources are listed in Table \ref{morph}.  The masses were computed by 
fitting the UV-NIR SEDs using {\it Le Phare} \citep{ilb10} and assuming a Chabrier \citep{cha03} initial mass function (IMF).  
The mass range is log~$(M/M_\sun) \sim 10.0 - 11.5$ corresponding to $\sim 0.5-3 M^*$.  Higher mass systems are more 
likely to be associated with higher infrared luminosity. 

\subsubsection{Spectroscopy: Extinction, Abundances and Spectral Types}
\begin{figure*} \centering
 
\includegraphics[scale=.8]{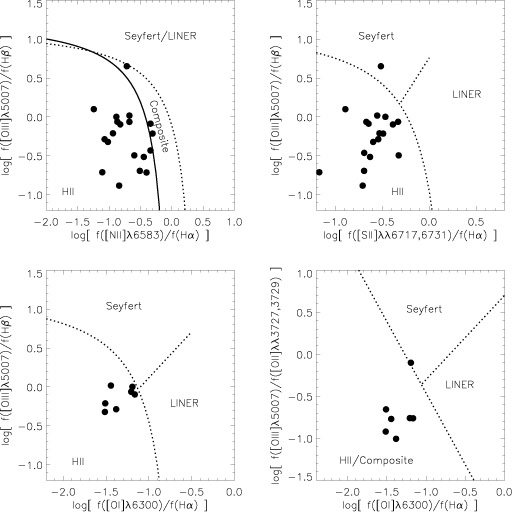}
\caption
{Optical spectral diagnostics, following the formalism of \citet{kew06}, for those sources with optical spectra and for which we were able to 
measure the relevant diagnostic lines.   The dotted lines divide the diagrams into spectral classification types, and the solid line denotes the limit to pure star formation. }
\label{stype}

\end{figure*}

Our high-resolution Keck/ESI spectra, supplemented by SDSS spectra and four low-resolution ESI spectra, allow us to 
measure robust spectral types for 25 of the 160\micron\ sources in our sample.  An example of these spectra was shown in 
Figure \ref{spec}.  Spectra for the 25 sources with spectral types as well as those from three sources for which we have 
data but were unable to measure spectral types are available in the online edition of the Journal.  After carefully accounting 
for the effects of stellar absorption and applying an extinction correction (median $E(B-V) = 0.7$), we classify the spectra 
as \ion{H}{2}-region-like, or star-forming (H); composite, or star formation + an AGN (C); Seyfert (S); or LINER (L), based on
the classification scheme proposed by \citet{kew06} (see Figure \ref{stype}).  Table \ref{lumtab} lists the measured 
spectral types and extinctions for each galaxy when they are available. 

\indent Dividing the subsample with spectral types into luminosity bins, we find that 15 of 19, (79\%), of galaxies in the
log$(L_{\rm IR}/L_\sun) < 11$ bin have  \ion{H}{2}-region-like spectral type, consistent with star formation as the dominant 
source of excitation.  This is close to the \ion{H}{2}-region-like fraction of nearby galaxies selected at 60\micron\ \citep[$\sim$70\% -- ][]{vei95, yua10}.  
The other four galaxies include one Seyfert and one LINER and 2 objects with mixed types that suggest  a ``composite" starburst-AGN 
mixture of excitation.  Only six galaxies with high resolution spectroscopy (and hence derived spectral types) 
have log$(L_{\rm IR}/L_\sun) > 11$.  Three (3/6 = 50\%) have \ion{H}{2}-region-like spectral type,  one is a Seyfert, one 
is a Seyfert/LINER, and one is a ``composite" mixture of starburst/LINER excitation.  Although the fraction of galaxies  with  
\ion{H}{2}-region-like spectral types decreases at higher infrared luminosity (similar to what is observed for nearby 
galaxies selected at 60\micron), the number statistics in this high luminosity bin are  too low to draw conclusions 
about the fractions of different spectral types.   In an attempt to provide additional information on the spectral types of 
our high infrared luminosity sources, we have employed a new technique developed from studies of SDSS galaxies \citep{smo08} that 
maps UV/optical continuum colors onto the spectral line diagnostic diagram.   This method is described in the Appendix,  
where the SDSS photometry for all of our  MIPS-160$\mu$m sources  is used to derive ``P1,P2" photometric 
spectral types for each source, following the prescription given 
by \citet{smo08}.  These results both confirm the large \ion{H}{2}-region-like fraction 
among the lower luminosity infrared sources, and show that composite and 
AGN spectral types appear to increase among  the highest luminosity sources.

\indent Because our spectra also contain the [\ion{O}{2}]
$\lambda\lambda$3727, 3729 doublet, we are able to estimate gas-phase
oxygen abundances for these systems.  Where available, these are
listed in Table \ref{morph}, using the robust
[\ion{N}{2}]/[\ion{O}{2}] diagnostic of \citet{kew02}.  To put these
in context, we also used the measured $K_{\rm s}$ data to compare to the
luminosity-metallicity relation in the NIR \citep{sal05}.  For the 8
systems that have sufficient information (upper-branch $R_{23}$ gas
abundances and measured luminosities; see \citet{rup08} for more on
the methodology), we find that 5 follow the $L-Z$ relation of normal
galaxies.  Three others have higher luminosities than the data
threshold, and appear to be slightly below the $L-Z$ relation (by
$0.1-0.2$~dex), as found for other infrared-selected objects at high
luminosity \citep{rup08}.

\subsubsection{Radio-FIR Correlation} 
\begin{figure*}[] 

\begin{center} 
\begin{tabular}{cc}
\includegraphics[scale=.5]{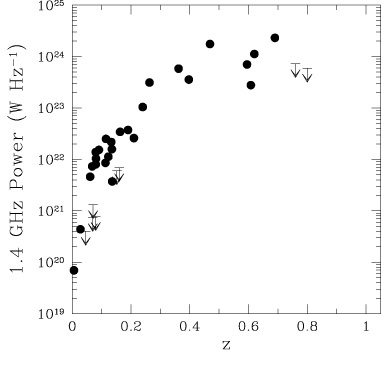}
\hspace{.1in}
\includegraphics[scale=.5]{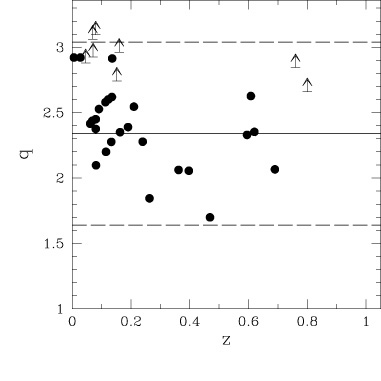}
\end{tabular}
\caption{Left panel: 1.4 GHz radio power of the 34 sources with available radio data shown as a function of their redshifts. 
Right panel: The radio-FIR correlation $q$-value of the same 34 sources are shown.  The solid horizontal line corresponds to $q=2.34$, which is the mean of the local star forming galaxies, while the lower and upper dashed lines mark the radio-excess and IR-excess objects, respectively \citep[see:][]{yun01b}.
}
\label{fig:radio}
\end{center}
\end{figure*}

The measured 1.4 GHz radio continuum fluxes are converted to 1.4 GHz radio power $L_{\rm 1.4GHz}$ assuming a spectral index of $\alpha =+0.75$\footnote{Spectral index $\alpha$ is defined as $S_\nu = S_0 (\frac{\nu}{\nu_0})^{-\alpha}$.}, and they are plotted as a function of redshift on the left panel of Figure \ref{fig:radio}.  The observed 1.4~GHz radio power range between $10^{20}$ and $10^{24.3}$ W Hz$^{-1}$ (see Table \ref{lumtab}), similar to the IR-selected galaxies in the local universe studied by \citet{yun01b}, and none of the sources has sufficient radio power to be classified as a ``radio-loud'' object.  The most luminous infrared sources also tend to be those with the largest radio luminosities, i.e. log$(L_{\rm 1.4GHz}) = 23.0-24.4$,  equivalent to the radio powers typically seen among Seyfert galaxies, and thus the presence of a low luminosity AGN cannot be ruled out based on these radio powers alone.

The well-known correlation between the measured infrared luminosity and radio power for star forming galaxies is often quantified using the ratio commonly referred to as ``$q$-value''
\begin{equation}q={\rm log}~[({\rm FIR}/3.75\times 10^{12}~{\rm Hz})/S_{1.4 {\rm GHz}}]\end{equation}
where FIR is the far-infrared flux density and $S_{1.4 {\rm GHz}}$ is in W m$^{-2}$ Hz$^{-1}$ \citep{condon92,yun01b}.  We computed these $q$-values using the $L_{\rm 1.4GHz}$ derived above and $L_{\rm FIR}$ computed from the best-fit SED models integrated between $\lambda=40$ and 500 \micron, where the wavelength range has been chosen to match the original definition of $L_{\rm FIR}$ used to compute ``$q$".  As shown on the right panel in Figure \ref{fig:radio}, the derived $q$-values of the LH 160 \micron\ sources fall between 1.6 and 3.0, suggesting that most of these sources follow the same radio-FIR correlation as the local star forming galaxy populations.  Some of the low redshift ($z\lesssim 0.15$) sources appear to have $q$-values on the high end of the local population.  These are also the sources with the largest angular size, and the VLA measurements are likely under-estimates as a consequence.  None of the LH 160 \micron\ sources has a $q$-value less than 1.6 and thus a clear evidence for a radio-loud AGN.

\section{Summary}

We have made use of Spitzer-SWIRE imaging data of two $\sim 0.5$deg$^2$ fields (LHEX and LHNW) in the Lockman Hole,
to  identify a complete sample of 40 MIPS-160$\mu$m selected extragalactic sources, with $S_{160} > 120$mJy.  
In combination with Keck spectroscopy and photometry from SDSS and 2MASS, we have obtained redshifts and 
infrared luminosities, and have attempted to characterize the host galaxy properties for all of the objects in the sample.  
The luminosity function for the MIPS-160$\mu$m sample has been compared with the ``local'' ($z \leq 0.05$) luminosity 
function of FIR galaxies previously derived using the IRAS all-sky survey.  
\\ 
\indent Our main results can be summarized as follows:
\\ 

\indent (1) The complete $S_{160} > 120$mJy sample contains 40 galaxies with infrared luminosities
in the range log$(L_{\rm IR}/L_{\sun}) = 8.81 - 12.74$, with a maximum redshift: $z=0.80$.  The cumulative 
source counts down to 120mJy are estimated to be $1.2 \times 10^5$ sources sr$^{-1}$ at 160$\mu$m. 
\\ 
\indent (2) The luminosity function of the sources with log$(L_{\rm IR}/L_{\sun}) \sim 9.5 - 11.5$ is similar to that found 
previously for infrared galaxies in the IRAS $60\mu$m local galaxy sample.
\\ 
\indent (3) The co-moving space density of the MIPS-160$\mu$m galaxies with log$(L_{\rm IR}/L_{\sun}) > 12$  
is $\sim 10 \times$ higher than that for the local infrared galaxies with similar infrared luminosities found in the 
IRAS RBGS.  Assuming pure number density evolution proportional to $(1+z)^n$, these results give $n=6 \pm 1$, 
which implies strong evolution of the most luminous infrared sources, in contrast to little or no evolution observed  
in the number density of lower luminosity objects with log$(L_{\rm IR}/L_{\sun}) < 11$. 
\\ 
\indent (4)  The host galaxy masses for our sample are in the range log$(M_\star/M_\sun)  \sim 10.0-11.5$ ($0.5-3 M^*$), with evidence for 
an increase in host mass from a mean of 10.7 to 11.0 for objects with infrared luminosities below and above log$(L_{\rm IR}/L_{\sun}) = 11$, respectively.
\\
\indent (5) The morphology and spectral types for our flux-limited sample of  160$\mu$m-selected sources 
generally agree with what has been observed locally for 60$\mu$m-selected samples.   At log$(L_{\rm IR}/L_{\sun}) > 11$, 
the fraction of disturbed and/or merger systems and the fraction of objects with ``composite" and/or AGN spectral types 
increases with increasing $L_{\rm IR}$.  At log$(L_{\rm IR}/L_{\sun}) < 11$ most objects appear to be either 
unperturbed spirals and/or weakly interacting systems with spectral types typical of HII regions.  
 \\
 \indent (6) None of the LH sources has sufficient radio power to be classified as a ``radio-loud'' object.  However, the most luminous infrared sources also tend to have the highest radio luminosities, i.e. log$(L_{\rm 1.4GHz}) = 23.0-24.4$, equivalent to the radio powers typically seen among Seyfert galaxies, and thus the presence of a low luminosity AGN cannot be ruled out based on these radio powers alone.
 \\ \\

\section{Acknowledgments}
We benefited from the published data and preliminary analyses of S. Oyabu.  VS acknowledges support from the Owens Valley Radio Observatory, which 
is supported by the National Science Foundation through grant AST-0838260, and also received funding from the
European Union's Seventh Framework programme under grant agreement 229517.  YT was financially supported in part by the 
Ministry of Education, Culture, Sports, Science and Technology (Nos. 10044052 and 10304013), and by the JSPS 
(Nos. 15340059, 17253001 and 19340046).  This research has made use of the 
NASA/IPAC Extragalactic Database (NED) which is operated by the Jet Propulsion Laboratory, California Institute of 
Technology, under contract with the National Aeronautics and Space Administration.   This publication makes use of 
data products from the Two Micron All Sky Survey, which is a joint project of the University of Massachusetts and the 
Infrared Processing and Analysis Center/California Institute of Technology, funded by the National Aeronautics and 
Space Administration and the National Science Foundation.  Funding for the SDSS and SDSS-II has been provided by 
the Alfred P. Sloan Foundation, the Participating Institutions, the National Science Foundation, the U.S. Department of 
Energy, the National Aeronautics and Space Administration, the Japanese Monbukagakusho, the Max Planck Society, 
and the Higher Education Funding Council for England. The SDSS Web Site is http://www.sdss.org/.

\indent The SDSS is managed by the Astrophysical Research Consortium for the Participating Institutions. The participating 
institutions are the American Museum of Natural History, Astrophysical Institute Potsdam, University of Basel, University of 
Cambridge, Case Western Reserve University, University of Chicago, Drexel University, Fermilab, the Institute for Advanced 
Study, the Japan Participation Group, Johns Hopkins University, the Joint Institute for Nuclear Astrophysics, the Kavli Institute 
for Particle Astrophysics and Cosmology, the Korean Scientist Group, the Chinese Academy of Sciences (LAMOST), 
Los Alamos National Laboratory, the Max-Planck-Institute for Astronomy (MPIA), the Max-Planck-Institute for Astrophysics (MPA), 
New Mexico State University, Ohio State University, University of Pittsburgh, University of Portsmouth, Princeton University, 
the United States Naval Observatory, and the University of Washington.

\appendix
\section{Using rest-frame colors to assign galaxy types}
\label{subsec:p1p2}
In an attempt to determine the spectral type of the 12 galaxies in our
sample for which we lack optical spectra, we make use of a photometric (rest-frame color based) method
extensively studied in \citet{smo08}.    \citet{smo06} have shown that
principal component rest-frame colors (P1 and P2 hereafter) drawn from
the modified Str\"{o}mgren photometric system (3500--5800~\AA ; \citealt{odell02})
essentially trace the position of a galaxy in the BPT diagram
\citep{bpt81,kewley01,kew06,kauffmann03}. Hence, they can be used as efficient tracers of galaxy
type, such as low-luminosity AGN (Seyfert and LINER), star forming, and
composite galaxies (see \citealt{smo06,smo08} for more details). Here
we extend the method developed by \citet{smo08}, that utilizes only
the P1 color to disentangle star forming from AGN galaxies, to a probabilistic approach that uses
both, P1 and P2 colors, and we adapt it to an IR- (rather than radio-)
selected sample.

\subsection{Derivation of P1, P2 colors}
Given that we have SDSS {\em ugriz} photometry for the complete MIPS-160$\mu$m  
sample, we derive the P1 and P2 colors for all 40 galaxies in our sample by
fitting their SEDs (encompassed by the SDSS {\em ugriz} photometry) 
with 100,000 spectra from the \citet{bc}
stellar population synthesis model library. Before performing the
$\chi^2$ minimization fit we redshift all the model spectra to the
galaxy's spectroscopic redshift.  The colors are then computed from
the best fit model spectrum (see Sec.~4.2.\ in \citealt{smo08} for
more details about the SED fitting).

\indent To assess the accuracy of photometrically synthesized colors for
IR-selected galaxies, we have derived the (P1,P2) colors via SED
fitting (in the same way as described above) for an {\em IR-selected
  control sample} with available (P1,P2) colors computed
independently from their spectra. The control sample, limited to a
redshift range of 0.04 to 0.3, contains $\sim1350$ galaxies drawn from
the SDSS DR1 ``main'' spectroscopic galaxy sample matched to the IRAS
Faint Source Catalog (see \citealt{obric06} for details about the
cross-correlation of the catalogs). The spectroscopically derived
(P1,P2) colors have been synthesized by \citet{smo06} by convolving
the SDSS high-resolution spectra with the Str\"{o}mgren filter system.
They estimated that these spectroscopically derived rest-frame colors
are accurate to $0.03$~mag.

\indent In \f{fig:p1p2offset} \ we show the difference between the
photometrically and spectroscopically derived (P1,P2) colors for our
IR-selected control galaxies as a function of the photometrically
derived P1 color. We use the median offset as a function of P1 to
correct the photometrically derived colors, i.e.\ to scale these to
the SDSS spectroscopic system. The distribution of the (P1,P2) color
differences after the corrections have been applied are shown in
\f{fig:p1p2histo} . As expected, the corrections have eliminated
systematic effects. Furthermore, as the accuracy of the
spectroscopically derived colors has been shown to be $0.03$~mag
\citep{smo06}, the accuracy of the photometrically derived
(P1,P2) colors is likely better than $0.12$ and $0.03$, 
respectively. It is remarkable that the accuracy of the P2 color
derived via SED fitting is comparable to that of the spectroscopically
derived color.

\subsection{Classifying the Lockman galaxies using the P1-P2 color
  method} 

In \f{fig:p1p2} \ we show the P1 vs.\ P2 the distribution for our 40
galaxies in our MIPS-160$\mu$m sample (their P1,P2 colors were
corrected for systematic offsets as described in the previous section;
filled dots). To assess the nature of these 40 galaxies for each one
we compute the probability that it is a star forming, composite, AGN
or absorption galaxy given its (P1,P2) rest-frame colors. The
probability is computed based on the underlying distribution of our
$\sim1300$ control (SDSS-IRAS; $0.04<z<0.3$) galaxies
(spectroscopically divided into absorption, star forming, AGN and
composite galaxies using the standard diagnostics;
\citealt{bpt81,kewley01,kauffmann03}) in the P1-P2 plane as follows. We
bin the P1-P2 plane in 2 dimensions. The size of the bins is taken to
be about 2 times the photometric color uncertainty (see previous
section). For each (P1,P2) bin we then calculate the probability as
the ratio of the number of each (spectroscopically classified) galaxy
type relative to the total number of control galaxies in that
bin. Given the photometrically synthesized (P1,P2) colors for our
Lockman galaxies we can then access the probability of each galaxy
being a star forming, AGN, composite or absorption galaxy. These
probability contours are shown in \f{fig:p1p2} , and our results are
summarized in Table \ref{spectab}.

 
\clearpage

 \begin{figure*}[] 

\begin{center}
\includegraphics[scale=.9]{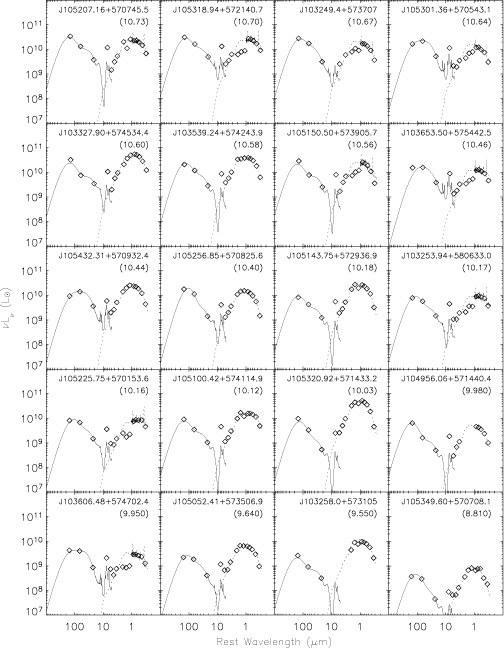}

\end{center}
Figure~\ref{seds1} (Continued).

\end{figure*}

\begin{figure*}
\begin{center}
\begin{tabular}{cccc}

\scriptsize J105207.16$+$570745.5, (10.73)&\scriptsize J105318.94$+$572140.7, (10.70)&\scriptsize J103249.4$+$573707, (10.67)&\scriptsize J105301.36$+$570543.1, (10.64)\\
\includegraphics*[scale=.65]{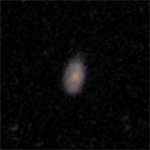} &
\includegraphics*[scale=.65]{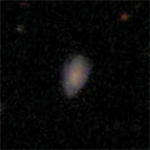} &
\includegraphics*[scale=.65]{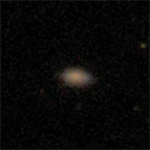} &
\includegraphics*[scale=.65]{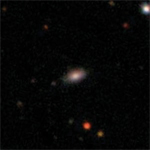} \\
\scriptsize J103327.90$+$574534.4, (10.60)&\scriptsize J103539.24$+$574243.9, (10.58)&\scriptsize J105150.50$+$573905.7, (10.56)&\scriptsize J103653.50$+$575442.5, (10.46)\\
\includegraphics*[scale=.65]{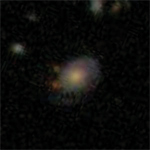} &
\includegraphics*[scale=.65]{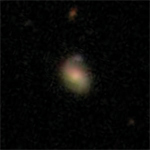} &
\includegraphics*[scale=.65]{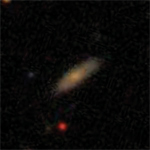} &
\includegraphics*[scale=.65]{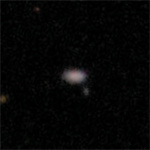} \\
\scriptsize J105432.31$+$570932.4, (10.44)&\scriptsize J105256.85$+$570825.6, (10.40)
&\scriptsize J105143.75$+$572936.9, (10.18)&\scriptsize J103253.94$+$580633.0, (10.17)\\
\includegraphics*[scale=.65]{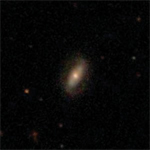} &
\includegraphics*[scale=.65]{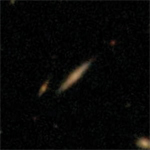} &
\includegraphics*[scale=.65]{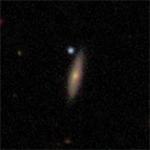} &
\includegraphics*[scale=.65]{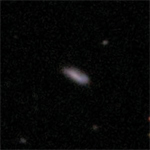} \\
\scriptsize J105225.75$+$570153.6, (10.16)&\scriptsize J105100.42$+$574114.9, (10.12)
&\scriptsize J105320.92$+$571433.2, (10.03)&\scriptsize J104956.06$+$571440.4, (9.98)\\
\includegraphics*[scale=.65]{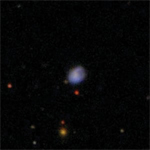} &
\includegraphics*[scale=.65]{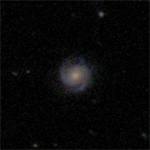} &
\includegraphics*[scale=.65]{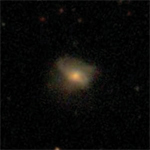} &
\includegraphics*[scale=.65]{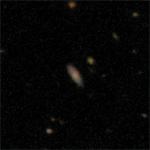} \\
\scriptsize J103606.48$+$574702.4, (9.95)&\scriptsize J105052.41$+$573506.9, (9.64)
&\scriptsize J103258.0$+$573105, (9.55)&\scriptsize J105349.60$+$570708.1, (8.81)\\
\includegraphics*[scale=.65]{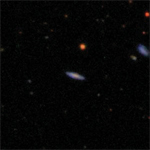} &
\includegraphics*[scale=.65]{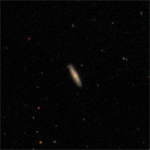} &
\includegraphics*[scale=.65]{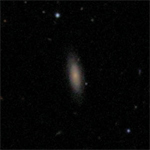} &
\includegraphics*[scale=.65]{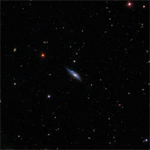} \\
\end{tabular}
\end{center}
Figure~\ref{grij} (Continued).

\end{figure*}

\begin{figure}
\center{
\includegraphics[scale=0.7]{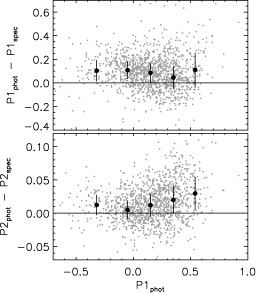}
\caption{ Offsets between the photometrically and spectroscopically
  derived P1 and P2 colors as a function of the former for a sample of
  $\sim1350$ SDSS-IRAS galaxies (small gray dots). The former were
  obtained via SED fitting to the SDSS $ugriz$ photometry using
  100,000 model spectra from the \citet{bc} library. The latter were
  derived from SDSS spectra as described in \citet{smo06}. The large
  dots represent median offsets as a function of the photometrically
  derived P1 color. The error bars show the interquartile ranges.
  \label{fig:p1p2offset}}
}
\end{figure}

\begin{figure}
\center{
\includegraphics[scale=0.85]{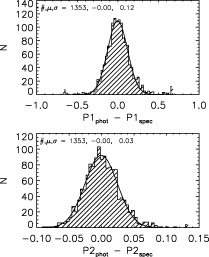}
\caption{ Distribution of the photometrically and spectroscopically
  derived P1 and P2 colors after the systematic offset correction (shown in
  \f{fig:p1p2offset} )  was applied. The number of
  objects, mean, and standard deviation, as well as the best Gaussian
  fit are also shown in the panels.
  \label{fig:p1p2histo}}
}
\end{figure}

\begin{figure}
\center{
\includegraphics[scale=.8]{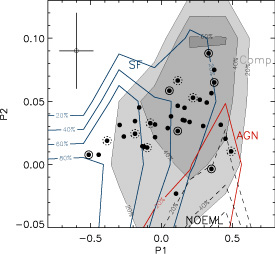} 
\caption{ P1 vs.\ P2 color plane. The colors for the Lockman galaxies
  (after the corrections for systematics have been applied) are shown
  by large dots. Sources with $11<\mathrm{\log{L_{IR}/L_\odot}}<12$
  and $\mathrm{\log{L_{IR}/L_\odot}}>12$ are encircled with dashed
  and full lines, respectively. Contours represent the absorption
  (i.e.\ galaxies with no emission lines; NOEML), star forming,
  composite and AGN probability
  levels given a (P1,P2) combination (see text for details).
The uncertainties in the P1 and P2 colors obtained
  via SED fitting are shown in the top left of the panel (see also
  \f{fig:p1p2histo} . Note, however, that the P2 error is likely lower
  than indicated in the panel (see text for details).
  \label{fig:p1p2}}
}
\end{figure}

\clearpage

\begin{landscape}
\begin{deluxetable}{rlcccccccccr} 
\tabletypesize{\scriptsize}
\setlength{\tabcolsep}{0.05in}
 \tablecaption{Source List and Catalog Cross-Identification}
 \tablewidth{0pc}
\tablehead{
   \colhead{\#}&\colhead{Source}&\colhead{RA}&\colhead{DEC}&\colhead{160$\micron$}&\colhead{70$\micron$}&\colhead{24$\micron$}&\colhead{IRAC+24$\micron$}&\colhead{NIR}&\colhead{Optical}&\colhead{C\_90}&\colhead{C\_160} \\
  \cline{5-8} \\
   &\colhead{(SWIRE ID\tablenotemark{1})}&\colhead{(J2000)}&\colhead{(J2000)}&\multicolumn{4}{c}{(SWIRE IDs)}&\colhead{(2MASS ID\tablenotemark{2})}&\colhead{(SDSS ID)}&\multicolumn{2}{c}{(ISO ID)}
   }
   \startdata

1&SWIRE3\_J103237.44$+$580845.9&10 32 37.45&58 08 46.0&797&1989&\nodata&620978&PSC:10323767+5808446&587729388220776605&\nodata&\nodata \\
2&SWIRE2\_24\_J103249.4$+$573707&10 32 49.48&57 37 07.8&938&2338&3313&\nodata&PSC:10324949+5737081&587729387683840107&1NW192&2NW003\\
3&SWIRE3\_J103253.94$+$580633.0&10 32 53.95&58 06 33.1&804&2006&\nodata&620883&PSC:10325395+5806333&587729388220776539&\nodata&\nodata \\
4&SWIRE2\_24\_J103258.0$+$573105&10 32 58.02&57 31 05.3&974&\nodata&2395&\nodata&XSC:10325794+5731068&587732582590513271&\nodata&\nodata \\
5&SWIRE2\_24\_J103314.8$+$573110&10 33 14.88&57 31 10.5&964&2385&2684&\nodata&\nodata&587732582590513949&\nodata&\nodata \\
6&SWIRE3\_J103320.32$+$574913.6&10 33 20.33&57 49 13.6&868&2177&\nodata&463679&\nodata&587732583127449867&1NW021&2NW005\\
7&SWIRE3\_J103327.90$+$574534.4&10 33 27.91&57 45 34.5&887&\nodata&\nodata&462097&XSC:10332785+5745351&587729387683905620&\nodata&\nodata \\
8&SWIRE3\_J103341.28$+$580221.4&10 33 41.29&58 02 21.4&805&2013&\nodata&472711&\nodata&587732583127515249&\nodata&2NW004\\
9&SWIRE3\_J103358.73$+$574317.1&10 33 58.73&57 43 17.2&891&2211&\nodata&462979&XSC:10335869+5743169&587729387683905627&1NW030&\nodata \\
10&SWIRE2\_24\_J103358.9$+$572952&10 33 58.92&57 29 52.3&945&2356&3540&\nodata&\nodata&587732582590578995&1NW272&2NW009\\
11&SWIRE3\_J103515.83$+$573337.4&10 35 15.84&57 33 37.5&903&2253&\nodata&462870&XSC:10351578+5733375&587732582590644271&\nodata&\nodata \\
12&SWIRE3\_J103526.79$+$575147.6&10 35 26.80&57 51 47.6&807&2034&\nodata&473946&\nodata&587729387683971243&\nodata&\nodata \\
13&SWIRE2\_24\_J103538.8$+$573546&10 35 38.87&57 35 46.1&884&2209&7870&\nodata&\nodata&587732582590644410&\nodata&\nodata \\
14&SWIRE3\_J103539.24$+$574243.9&10 35 39.24&57 42 43.9&851&2117&\nodata&469789&XSC:10353919+5742444&587732582590644435&\nodata&\nodata \\
15&SWIRE2\_24\_J103557.1$+$572234&10 35 57.12&57 22 34.1&941&2340&5113&\nodata&PSC:10355706+5722341&587729387147100345&\nodata&\nodata \\
16&SWIRE3\_J103603.97$+$574812.5&10 36 03.97&57 48 12.6&811&2038&\nodata&474542&\nodata&587729387684036627&1NW092&\nodata \\
17&SWIRE3\_J103606.48$+$574702.4&10 36 06.48&57 47 02.5&812&2052&\nodata&474042&PSC:10360650+5747024&587729387684036713&1NW023&\nodata \\
18&SWIRE3\_J103653.50$+$575442.5&10 36 53.50&57 54 42.6&773&1935&\nodata&481645&PSC:10365355+5754426&587729387684102266&\nodata&\nodata \\
19&SWIRE3\_J104948.86$+$573458.2&10 49 48.87&57 34 58.3&558&1482&\nodata&228325&XSC:10494884+5734579&587729386611015724&\nodata&\nodata \\
20&SWIRE3\_J104956.06$+$571440.4&10 49 56.07&57 14 40.5&652&\nodata&\nodata&216021&\nodata&587729386074079354&\nodata&\nodata \\
21&SWIRE3\_J105041.96$+$570706.7&10 50 41.96&57 07 06.8&673&1741&\nodata&215353&XSC:10504192+5707064&587732580980818115&1EX048&2EX004\\
22&SWIRE3\_J105052.41$+$573506.9&10 50 52.41&57 35 07.0&534&1401&\nodata&233199&XSC:10505236+5735064&587732581517754577&1EX041&2EX013\\
23&SWIRE3\_J105056.60$+$571631.2&10 50 56.60&57 16 31.2&625&1638&\nodata&222478&\nodata&587729386074145329&1EX085&2EX068\\
24&SWIRE3\_J105100.42$+$574114.9&10 51 00.43&57 41 15.0&512&1338&\nodata&237144&XSC:10510035+5741153&587729386611081359&\nodata&\nodata \\
25&SWIRE3\_J105113.41$+$571425.9&10 51 13.41&57 14 26.0&617&1642&\nodata&222633&\nodata&587729386074145391&1EX081&2EX115\\
26&SWIRE3\_J105143.75$+$572936.9&10 51 43.75&57 29 36.9&544&1420&\nodata&234071&XSC:10514374+5729367&587732581517754609&\nodata&\nodata \\
27&SWIRE3\_J105150.50$+$573905.7&10 51 50.51&57 39 05.7&499&\nodata&\nodata&239860&PSC:10515055+5739054&587729386611146842&\nodata&\nodata \\
28&SWIRE3\_J105151.64$+$570935.7&10 51 51.64&57 09 35.7&634&\nodata&\nodata&72224&\nodata&587732580980883908&1EX047&2EX036\\
29&SWIRE3\_J105207.16$+$570745.5&10 52 07.16&57 07 45.6&635&1667&\nodata&72439&XSC:10520715+5707445&587732580980883557&1EX034&2EX016\\
30&SWIRE3\_J105225.75$+$570153.6&10 52 25.76&57 01 53.7&653&1703&\nodata&70615&PSC:10522572+5701537&587732580980883560&\nodata&\nodata \\
31&SWIRE3\_J105242.40$+$572444.7&10 52 42.40&57 24 44.8&542&1413&\nodata&235900&PSC:10524240+5724447&587729386074276002&\nodata&\nodata \\
32&SWIRE3\_J105252.76$+$570753.7\tablenotemark{3}&10 52 52.77&57 07 53.8&\nodata&1607&\nodata&76811&XSC:10525283+5707537&587732580980949106&1EX028&\nodata \\
33&SWIRE3\_J105256.85$+$570825.6&10 52 56.85&57 08 25.7&603&1606&\nodata&77421&XSC:10525689+5708257&587732580980949119&1EX269&2EX047\\
34&SWIRE3\_J105301.36$+$570543.1&10 53 01.36&57 05 43.2&611&1629&\nodata&76215&PSC:10530133+5705433&587732580980949116&1EX062&\nodata \\
35&SWIRE3\_J105308.32$+$570645.6&10 53 08.32&57 06 45.6&610&\nodata&\nodata&77408&\nodata&587732580980949913&\nodata&\nodata \\
36&SWIRE3\_J105314.84$+$574137.6&10 53 14.85&57 41 37.7&447&1202&\nodata&247971&\nodata&587732581517885646&\nodata&\nodata \\
37&SWIRE3\_J105318.94$+$572140.7&10 53 18.94&57 21 40.7&538&1407&\nodata&86339&PSC:10531899+5721394&587729386074276027&1EX179&\nodata \\
38&SWIRE3\_J105320.92$+$571433.2&10 53 20.92&57 14 33.3&568&1498&\nodata&82519&XSC:10532085+5714338&587732580980949157&1EX126&\nodata \\
39&SWIRE3\_J105349.60$+$570708.1&10 53 49.60&57 07 08.1&582&1562&\nodata&80766&XSC:10534950+5707075&587732580980949076&\nodata&\nodata \\
40&SWIRE3\_J105432.31$+$570932.4&10 54 32.31&57 09 32.5&559&1478&\nodata&85419&XSC:10543226+5709324&587732580981014580&\nodata&\nodata \\  
 \enddata

  \tablecomments{Sources in the LHNW and LHEX fields.  The columns list a counter number, SWIRE Source ID, right ascension and declination (HH MM SS.SS and DD MM SS.S) taken from from the SWIRE IRAC+24$\micron$ when available, ID counter numbers from the SWIRE 160$\micron$, 70$\micron$, 24$\micron$, and IRAC+24$\micron$ catalogs, 2MASS source designation and catalog, object ID from SDSS, and the ISO-LH 90$\mu$m and 160$\micron$ source ID from \citet{oya05}.}

  \tablenotetext{1}{SWIRE2\_24\_J103249.4$+$573707, J103258.0$+$573105, J103314.8$+$573110, J103358.9$+$572952, J103538.8$+$573546, and J103557.1$+$572234 lack entries in the SWIRE IRAC+24$\micron$ catalog so their RA, DEC and 24$\micron$ flux are taken from the independent SWIRE 24$\micron$ catalog}
\tablenotetext{2}{Preceding the 2MASS source designation are the reference catalog abbreviations, PSC for the 2MASS Point Source Catalog, and XSC for the Extended Source Catalog.  Values from the XSC are preferred when available.} 

\tablenotetext{3}{SWIRE3\_J105252.76$+$570753.7 was not included in the SWIRE 160\micron\ catalog, but was visible in the image, and detected by ISO so we independently measure its flux by hand.}
  \label{counter}

\end{deluxetable}
\clearpage
\end{landscape}

\begin{landscape}
\begin{deluxetable}{rlccrrrrrrcccrrrrrrrr}
\thispagestyle{empty}
\tabletypesize{\scriptsize}
\setlength{\tabcolsep}{0.05in}
\tablecaption{Source Photometry}
\tablewidth{0pt}

\tablehead{
\colhead{\#}&\colhead{Source}&\colhead{20cm}&\colhead{160$\mu$m}&\colhead{70$\mu$m}&\colhead{24$\mu$m}&\colhead{8$\mu$m}&\colhead{5.8$\mu$m}&\colhead{4.5$\mu$m}&\colhead{3.6$\mu$m}&\colhead{$K\! s$}&\colhead{$H$}&\colhead{$J$}&\colhead{$i'$}&\colhead{$r'$}&\colhead{$g'$}&\colhead{$z$}&\colhead{$i$}&\colhead{$r$}&\colhead{$g$}&\colhead{$u$} \\
\cline{3-10} & \cline{10-20} \\
&\colhead{(SWIRE ID)}&\multicolumn{8}{c}{(mJy)}&\multicolumn{11}{c}{(mag)}
}

\startdata
1&SWIRE3\_J103237.44$+$580845.9&\nodata&390&117&6.17&2.051&0.409&0.437&0.401&17.13&17.09&18.32&\nodata&\nodata&\nodata&18.02&18.43&18.95&20.07&21.33\\
2&SWIRE2\_24\_J103249.4$+$573707&0.638&193&52.7&3.26&\nodata&0.641&\nodata&0.931&16.72&16.67&17.17&\nodata&\nodata&\nodata&16.79&17.05&17.47&18.22&19.57\\
3&SWIRE3\_J103253.94$+$580633.0&\nodata&142&45.1&3.96&4.067&0.697&0.550&0.785&17.12&16.82&17.15&\nodata&\nodata&\nodata&16.44&16.61&16.93&17.39&18.50\\
4&SWIRE2\_24\_J103258.0$+$573105&$<0.078$&124&18.6&2.23&\nodata&\nodata&\nodata&\nodata&15.27&14.80&15.27&\nodata&\nodata&\nodata&15.32&15.61&15.99&16.65&18.10\\
5&SWIRE2\_24\_J103314.8$+$573110&$<0.074$&150&18.9&0.84&\nodata&\nodata&\nodata&\nodata&-----&-----&-----&\nodata&\nodata&\nodata&20.55&21.22&22.44&23.37&23.07\\
6&SWIRE3\_J103320.32$+$574913.6&0.437&176&50.3&3.65&1.899&0.243&0.357&0.324&-----&-----&-----&18.53&18.97&19.83&18.45&18.71&19.16&19.90&20.85\\
7&SWIRE3\_J103327.90$+$574534.4&$<0.079$&122&12.5&1.97&2.014&0.272&0.610&0.800&16.31&16.01&15.98&16.88&17.12&17.90&16.20&16.48&16.88&17.65&18.91\\
8&SWIRE3\_J103341.28$+$580221.4&0.315&242&108&4.79&2.976&0.292&0.515&0.567&-----&-----&-----&17.90&18.35&18.88&20.99&21.66&21.16&22.70&22.38\\
9&SWIRE3\_J103358.73$+$574317.1&1.027&144&45.9&4.18&2.093&0.572&0.942&1.088&15.44&16.01&16.24&17.03&17.48&18.61&16.48&16.89&17.38&18.46&19.70\\
10&SWIRE2\_24\_J103358.9$+$572952&1.146&185&55.7&1.83&\nodata&0.088&\nodata&0.124&-----&-----&-----&\nodata&\nodata&\nodata&19.69&20.36&21.00&22.38&24.18\\
11&SWIRE3\_J103515.83$+$573337.4&$<0.080$&153&25.3&2.42&2.875&0.879&1.144&1.714&15.20&15.25&15.52&16.88&17.15&17.95&16.81&16.36&16.71&17.61&20.05\\
12&SWIRE3\_J103526.79$+$575147.6&0.224&139&26.3&2.01&0.183&0.155&0.111&0.159&-----&-----&-----&19.85&20.38&21.57&19.67&20.13&20.86&22.12&23.82\\
13&SWIRE2\_24\_J103538.8$+$573546&0.465&199&62.9&14.8&\nodata&\nodata&\nodata&\nodata&-----&-----&-----&\nodata&\nodata&\nodata&17.20&17.32&18.38&19.23&21.51\\
14&SWIRE3\_J103539.24$+$574243.9&0.227&150&37.4&3.04&3.130&1.001&1.171&1.645&15.12&15.36&15.57&16.05&16.46&17.30&15.85&16.16&16.59&17.43&18.94\\
15&SWIRE2\_24\_J103557.1$+$572234&\nodata&156&24.4&2.41&\nodata&0.600&\nodata&1.129&16.73&16.59&17.16&\nodata&\nodata&\nodata&16.88&17.34&17.95&19.19&20.88\\
16&SWIRE3\_J103603.97$+$574812.5&\nodata&172&75.5&4.26&0.507&0.210&0.147&0.136&-----&-----&-----&19.42&19.74&20.30&19.36&19.42&19.68&20.24&20.51\\
17&SWIRE3\_J103606.48$+$574702.4&\nodata&220&91.3&6.65&5.419&1.384&0.619&0.898&16.90&17.27&17.39&16.79&17.03&17.44&16.53&16.76&17.08&17.51&18.53\\
18&SWIRE3\_J103653.50$+$575442.5&\nodata&137&62.3&5.17&3.985&0.434&0.455&0.664&17.35&17.30&17.46&17.12&17.41&17.98&16.88&17.06&17.42&17.96&19.03\\
19&SWIRE3\_J104948.86$+$573458.2&0.276&242&45.9&3.77&3.756&0.730&0.915&1.200&15.89&15.53&16.09&17.15&17.63&18.52&16.48&16.88&17.42&18.46&20.21\\
20&SWIRE3\_J104956.06$+$571440.4&$<0.058$&126&14.0&1.48&1.830&0.213&0.374&0.501&-----&-----&-----&17.43&17.78&18.49&17.08&17.38&17.79&18.54&19.84\\
21&SWIRE3\_J105041.96$+$570706.7&0.669&362&75.2&5.30&7.510&1.552&1.587&2.330&14.84&14.84&15.36&16.10&16.46&17.37&15.77&16.15&16.64&17.59&19.44\\
22&SWIRE3\_J105052.41$+$573506.9&0.245&288&106&7.97&7.632&3.171&2.581&4.199&14.22&14.02&14.35&14.99&15.32&16.11&14.69&15.03&15.45&16.24&17.83\\
23&SWIRE3\_J105056.60$+$571631.2&0.379&124&39.8&2.10&0.639&0.142&0.131&0.124&-----&-----&-----&20.15&20.37&21.71&19.77&20.20&20.53&21.73&22.00\\
24&SWIRE3\_J105100.42$+$574114.9&$<0.100$&174&29.1&3.00&4.375&0.795&0.879&1.298&15.40&15.09&15.75&15.88&16.11&16.66&15.78&15.99&16.31&16.87&18.14\\
25&SWIRE3\_J105113.41$+$571425.9&0.802&250&37.4&2.95&0.313&0.295&0.370&0.414&-----&-----&-----&19.48&20.09&21.71&19.03&19.51&20.15&21.84&26.13\\
26&SWIRE3\_J105143.75$+$572936.9&0.588&121&26.9&2.68&4.030&1.009&1.146&1.725&15.05&14.86&15.53&16.03&16.47&17.15&15.52&15.86&16.34&17.16&18.89\\
27&SWIRE3\_J105150.50$+$573905.7&0.064&139&16.4&1.82&1.918&0.296&0.531&0.747&17.19&17.13&17.25&16.94&17.34&18.22&16.73&17.07&17.53&18.41&19.99\\
28&SWIRE3\_J105151.64$+$570935.7&0.083&140&16.1&1.06&0.106&\nodata&0.078&0.100&-----&-----&-----&20.40&20.90&22.15&20.32&20.53&21.10&22.42&22.68\\
29&SWIRE3\_J105207.16$+$570745.5&0.246&202&34.6&3.37&3.563&0.314&0.524&0.715&15.87&16.81&16.20&\nodata&\nodata&\nodata&16.66&16.86&17.24&17.86&19.06\\
30&SWIRE3\_J105225.75$+$570153.6&0.490&213&77.3&5.85&4.769&0.901&0.615&0.885&16.52&17.22&17.24&\nodata&\nodata&\nodata&16.24&16.30&16.55&16.86&17.87\\
31&SWIRE3\_J105242.40$+$572444.7&0.283&147&41.3&3.47&3.711&0.517&0.420&0.625&16.98&16.59&17.23&\nodata&\nodata&\nodata&16.79&16.94&17.26&17.78&18.90\\
32&SWIRE3\_J105252.76$+$570753.7&0.381&149&45.4&4.34&2.048&0.399&0.611&0.781&15.86&16.05&16.31&\nodata&\nodata&\nodata&16.69&16.99&17.47&18.36&19.84\\
33&SWIRE3\_J105256.85$+$570825.6&0.467&266&74.1&3.15&3.070&0.702&0.847&1.212&15.54&15.51&15.77&\nodata&\nodata&\nodata&16.17&16.55&17.03&17.92&19.73\\
34&SWIRE3\_J105301.36$+$570543.1&0.809&249&142&11.7&8.517&1.159&0.817&1.176&16.51&16.32&16.48&\nodata&\nodata&\nodata&16.35&16.54&17.00&17.59&18.93\\
35&SWIRE3\_J105308.32$+$570645.6&$<0.107$&142&15.5&0.51&0.072&0.092&0.084&0.118&-----&-----&-----&\nodata&\nodata&\nodata&20.48&21.16&22.68&23.15&24.97\\
36&SWIRE3\_J105314.84$+$574137.6&0.153&131&39.1&1.58&0.810&0.140&0.189&0.232&-----&-----&-----&18.11&18.34&19.07&18.05&18.18&18.52&19.24&20.19\\
37&SWIRE3\_J105318.94$+$572140.7&0.394&156&36.1&3.75&3.338&0.464&0.505&0.713&17.33&17.35&17.55&\nodata&\nodata&\nodata&16.64&16.91&17.27&17.88&18.91\\
38&SWIRE3\_J105320.92$+$571433.2&$<0.045$&144&21.6&1.19&1.863&1.392&2.110&3.230&14.37&14.30&14.72&\nodata&\nodata&\nodata&14.75&15.13&15.61&16.59&18.50\\
39&SWIRE3\_J105349.60$+$570708.1&0.911&1087&379&19.8&12.89&6.983&5.458&8.529&13.52&13.13&13.21&\nodata&\nodata&\nodata&13.76&13.83&14.03&15.28&16.25\\
40&SWIRE3\_J105432.31$+$570932.4&0.614&195&129&11.4&6.200&1.518&1.569&2.426&14.90&14.82&14.84&\nodata&\nodata&\nodata&15.23&15.52&15.93&16.69&18.15\\
\enddata
\thispagestyle{empty}
\tablecomments{Source photometry in the LHNW and LHEX fields listed in order of right ascension.  The columns list a counter number, the source ID, the 20 cm VLA flux, the flux in the Spitzer wavelengths in mJy, and 2MASS $JHK_{\rm s}$-bands, SWIRE $i'r'g'$-bands and SDSS $zirgu$-bands in AB magnitudes.  Sources without listed values (...) in Spitzer-IRAC and/or the SWIRE/NOAO $i'r'g'$-bands were typically located on the edge of the LHNW field not covered by SWIRE (see Figure \ref{cov}).   Sources without 2MASS photometry (-----) were not found in the 2MASS photometry catalog. }

\label{fltab}
\end{deluxetable}
\clearpage
\end{landscape}

\begin{deluxetable}{rllclr}
\tabletypesize{\scriptsize}
\tablecaption{Redshift and Infrared Luminosity}
\tablewidth{0pt}

\tablehead{
\colhead{\#}&\colhead{Source}&\colhead{$z$}&\colhead{$z$-Ref.}&\colhead{log$(L_{\rm IR})$}&\colhead{log$(L_{\rm 1.4GHz})$}\\
\multicolumn{4}{c}{} & \colhead{$(L_{\sun})$} & \colhead{(W Hz$^{-1}$)}
}
\startdata
1&SWIRE3\_J103237.44$+$580845.9\tablenotemark{*}&$0.42\pm0.02$&Phot-z&$12.28\pm0.05$&\nodata \\
2&SWIRE2\_24\_J103249.4$+$573707&0.115&ESI/ech&10.67&22.40\\
3&SWIRE3\_J103253.94$+$580633.0&0.073&SDSS&10.17&\nodata \\
4&SWIRE2\_24\_J103258.0$+$573105&0.046&SDSS&9.55&$<20.60$\\
5&SWIRE2\_24\_J103314.8$+$573110\tablenotemark{*}&$0.80^{+0.22}_{-0.18}$&Phot-z&$12.48^{+0.26}_{-0.27}$&$<23.77$\\
6&SWIRE3\_J103320.32$+$574913.6&0.240&ESI/ech&11.37&23.02\\
7&SWIRE3\_J103327.90$+$574534.4&0.152&SDSS&10.60&$<21.78$\\
8&SWIRE3\_J103341.28$+$580221.4\tablenotemark{*}&$0.62^{+0.06}_{-0.08}$&Phot-z\tablenotemark{2}&$12.62^{+0.10}_{-0.15}$ &24.05\\
9&SWIRE3\_J103358.73$+$574317.1&0.263&ESI/ech&11.46&23.49\\
10&SWIRE2\_24\_J103358.9$+$572952&0.469&ESI/ech&12.04&24.24\\
11&SWIRE3\_J103515.83$+$573337.4\tablenotemark{*}&$0.16\pm0.02$&Phot-z&$10.87\pm0.12$ &$<21.84$\\
12&SWIRE3\_J103526.79$+$575147.6&0.595&ESI/low-d&12.25 &23.85\\
13&SWIRE2\_24\_J103538.8$+$573546\tablenotemark{*}&$0.69\pm0.03$&Phot-z&$12.66^{+0.05}_{-0.04}$&24.36\\
14&SWIRE3\_J103539.24$+$574243.9&0.113&ESI/ech&10.58&21.93\\
15&SWIRE2\_24\_J103557.1$+$572234\tablenotemark{*}&$0.46^{+0.02}_{-0.26}$&Phot-z&$11.92^{+0.04}_{-0.84}$&\nodata \\
16&SWIRE3\_J103603.97$+$574812.5\tablenotemark{1}&0.511&B\tablenotemark{1}&12.20&\nodata \\
17&SWIRE3\_J103606.48$+$574702.4&0.044&SDSS&9.95&\nodata \\
18&SWIRE3\_J103653.50$+$575442.5&0.102&SDSS&10.46&\nodata \\
19&SWIRE3\_J104948.86$+$573458.2&0.135&SDSS+ESI/low-d&10.89&22.20\\
20&SWIRE3\_J104956.06$+$571440.4\tablenotemark{*}&$0.07\pm0.03$&Phot-z&$9.98^{+0.37}_{-0.46}$&$<20.87$\\
21&SWIRE3\_J105041.96$+$570706.7&0.091&ESI/ech&10.76&22.18\\
22&SWIRE3\_J105052.41$+$573506.9&0.028&SDSS+ESI/ech&9.64&20.64\\
23&SWIRE3\_J105056.60$+$571631.2&0.397&ESI/low-d&11.69&23.55\\
24&SWIRE3\_J105100.42$+$574114.9&0.071&ESI/ech&10.12&$<21.12$\\
25&SWIRE3\_J105113.41$+$571425.9&0.362&ESI/ech&11.90 &23.76\\
26&SWIRE3\_J105143.75$+$572936.9&0.081&ESI/ech&10.18 &22.01\\
27&SWIRE3\_J105150.50$+$573905.7&0.136&SDSS&10.56&21.57\\
28&SWIRE3\_J105151.64$+$570935.7&0.608&ESI/low-d&12.12 &23.44\\
29&SWIRE3\_J105207.16$+$570745.5&0.123&ESI/ech&10.73&22.05\\
30&SWIRE3\_J105225.75$+$570153.6&0.061&ESI/ech&10.16&21.66\\
31&SWIRE3\_J105242.40$+$572444.7\tablenotemark{*}&$0.19^{+0.01}_{-0.03}$&Phot-z&$11.05^{+0.07}_{-0.15}$&22.57\\
32&SWIRE3\_J105252.76$+$570753.7&0.163&SDSS+ESI/low-d&10.99&22.53\\
33&SWIRE3\_J105256.85$+$570825.6&0.080&ESI/low-d&10.40 &21.90\\
34&SWIRE3\_J105301.36$+$570543.1&0.080&ESI/ech&10.64&22.14\\
35&SWIRE3\_J105308.32$+$570645.6\tablenotemark{*}&$0.76^{+0.14}_{-0.26}$&Phot-z&$12.74^{+0.19}_{-0.44}$&$<23.86$\\
36&SWIRE3\_J105314.84$+$574137.6\tablenotemark{*}&$0.21^{+0.05}_{-0.03}$&Phot-z&$11.04^{+0.21}_{-0.15}$&22.41\\
37&SWIRE3\_J105318.94$+$572140.7&0.133&SDSS&10.70&22.34\\
38&SWIRE3\_J105320.92$+$571433.2&0.080&SDSS&10.03&$<20.89$\\
39&SWIRE3\_J105349.60$+$570708.1\tablenotemark{1}&0.006&NED\tablenotemark{1}&8.81&19.84\\
40&SWIRE3\_J105432.31$+$570932.4&0.068&SDSS&10.44 &21.86\\
\enddata

\tablecomments{Sources in the LHNW and LHEX fields listed in order of
  increasing right ascension, with columns giving a counter number, source ID, redshift,
  infrared luminosity (8-1000$\micron$) in units of $L_{\sun}$, and radio power $L_{1.4 GHz}$ in units: log(W Hz$^{-1}$).
  The sources of the spectra listed in the $z$-Ref. column are: from ESI
  on Keck II in low-dispersion (low-d) and echellette (ech) mode, from
  Amy Barger's private communication (2003) (B), from SDSS, and from NED.}
\tablenotetext{*}{Sources without existing optical spectra, for which we have determined photometric redshifts (see text and Figure \ref{seds1}).  The 
computed phot-z value, and 68\% uncertainty is listed.  This uncertainty was propagated when computing the listed infrared luminosity. }
\tablenotetext{1}{Thirty of the forty sources in our sample have spectroscopic redshifts, and 28 of these have spectra, which are displayed in Figures \ref{spec}.1 -- \ref{spec}.28.  The spectra for sources J103603.97$+$574812.5, and J105349.60$+$570708.1 are not available.}
\tablenotetext{2}{For the case of J103341.28+580221.4, we prefer the SWIRE optical data due to confusion with a foreground spiral and use these rather than the SDSS data to calculate its photometric redshift.}  

\label{lumtab}
\end{deluxetable}

\clearpage

\begin{deluxetable}{ccccccc}
\tablecaption{Luminosity Function}

\tablewidth{0pt}
\tablehead{
    \colhead{log$(L_{\rm IR})$}&  \colhead{$\Phi$} & \colhead{$\sigma_{\Phi}$} &  \colhead{$V/V_{\rm max}$}&  \colhead{$\sigma_{V/V_{\rm max}}$}& \colhead{Number} & \colhead{Median $z$}\\
\colhead{$(L_{\sun})$}& \colhead{(Mpc$^{-3}$ mag$^{-1}$)} & \multicolumn{5}{c}{}
}
    \startdata
9.8&$3.1\times 10^{-3}$&$1.8\times 10^{-3}$&0.53&0.11&3&0.044\\
10.2&$1.4\times 10^{-3}$&$5.6\times 10^{-4}$&0.64&0.09&6&0.077\\
10.6&$7.2\times 10^{-4}$&$2.3\times 10^{-4}$&0.59&0.07&10&0.114\\
11.0&$8.3\times 10^{-5}$&$3.7\times 10^{-5}$&0.68&0.11&5&0.163\\
11.4&$8.7\times 10^{-6}$&$6.1\times 10^{-6}$&0.66&0.17&2&0.251\\
11.8&$2.4\times 10^{-6}$&$1.4\times 10^{-6}$&0.65&0.13&3&0.397\\
12.2&$1.4\times 10^{-6}$&$6.2\times 10^{-7}$&0.57&0.09&5&0.511\\
12.6&$3.6\times 10^{-7}$&$1.8\times 10^{-7}$&0.58&0.10&4&0.725\\
\enddata
  \tablecomments{The galaxy sample is placed into luminosity bins of 0.4 in the range log$(L_{\rm IR}/L_{\sun})$.  We list the space density $\Phi$, the uncertainty in this value, volume sampling parameter $V/V_{\rm max}$, uncertainty in $V/V_{\rm max}$, the number of galaxies in each bin, and the median redshift.}
  \label{lfparam}

\end{deluxetable}

\clearpage
\begin{landscape}
\begin{deluxetable}{rlcccccccl}
\thispagestyle{empty}
\tabletypesize{\scriptsize}

\tablecaption{Spectral Properties, Masses and Morphologies}
\tablewidth{0pt}
\setlength{\tabcolsep}{0.05in}
\tablehead{
\colhead{\#}&\colhead{Source}&\colhead{log$(L_{\rm IR})$}&\colhead{$E(B-V)$}&\colhead{12+log$[O/H]$}&\colhead{S-Type}&\colhead{{\it q}-value}&\colhead{log$(M_{*})$}&\colhead{Morphology}&\colhead{Notes}\\
\multicolumn{2}{c}{}&\colhead{($L_{\sun}$)}&\multicolumn{4}{c}{}&\colhead{$(M_{\sun})$}&\multicolumn{2}{c}{}
}
\startdata
35&SWIRE3\_J105308.32$+$570645.6&$12.74^{+0.19}_{-0.44}$&\nodata &\nodata&\nodata&$>2.84$&10.90&Highly Disturbed&tidal features\\
13&SWIRE2\_24\_J103538.8$+$573546&$12.66^{+0.05}_{-0.04}$&\nodata&\nodata&\nodata&2.07&\tablenotemark{*}&Spheroid&S overlapping foreground galaxy with $z = 0.103$.\\
8&SWIRE3\_J103341.28$+$580221.4&$12.62^{+0.10}_{-0.15}$&\nodata &\nodata&\nodata&2.35&\tablenotemark{*}&Spheroid&SW foreground spiral with $z = 0.075$.\\
5&SWIRE2\_24\_J103314.8$+$573110&$12.48^{+0.26}_{-0.27}$&\nodata&\nodata&\nodata&$>2.66$&11.25&Spheroid&\\
1&SWIRE3\_J103237.44$+$580845.9&$12.28\pm0.05$&\nodata&\nodata&\nodata&\nodata&11.18&Disk&disturbed, tidal features\\
12&SWIRE3\_J103526.79$+$575147.6&12.25&\nodata &\nodata &H/C/L:&2.33&10.82&Spheroid&pair ($d = 20$~kpc, $\Delta z = 0.001$); star to S. \\
16&SWIRE3\_J103603.97$+$574812.5&12.20&\nodata &\nodata&\nodata&\nodata&10.88&Merger&overlapping pair\\
28&SWIRE3\_J105151.64$+$570935.7&12.12&\nodata &\nodata &S:&2.63&\tablenotemark{*}&Spheroid&pair ? ($d$ = 30~kpc)\\
10&SWIRE2\_24\_J103358.9$+$572952&12.04&\nodata &\nodata&\nodata&1.70&10.75&Spheroid&tidal debris.  (NE object is bkg source with $z = 0.837$.)\\
15&SWIRE2\_24\_J103557.1$+$572234&$11.92^{+0.04}_{-0.84}$&\nodata&\nodata&\nodata&\nodata&11.77&Disk&large (100 kpc) edge-on disk\\
25&SWIRE3\_J105113.41$+$571425.9&11.90&\nodata &\nodata&S/L:&2.06&11.49&Spheroid&compact\\
23&SWIRE3\_J105056.60$+$571631.2&11.69&1.00&8.87&H:&2.06&10.63&Spheroid&tidal debris + pair ($d$ = 40~kpc, $\Delta z = 0.001$) \\
9&SWIRE3\_J103358.73$+$574317.1&11.46&0.88&\nodata&H&1.84&11.70&Spheroid&faint SE companion ($d$ = 7~kpc)\\
6&SWIRE3\_J103320.32$+$574913.6&11.37&0.66&8.89&H&2.28&10.77&Disk&tidal arm(s).\\
31&SWIRE3\_J105242.40$+$572444.7&$11.05^{+0.07}_{-0.15}$&\nodata&\nodata&\nodata&2.39&10.86&Disk&\\
36&SWIRE3\_J105314.84$+$574137.6&$11.04^{+0.21}_{-0.15}$&\nodata&\nodata&\nodata&2.55&10.46&Disk&interacting system with SE companion ($d$ = 30~kpc)\\
32&SWIRE3\_J105252.76$+$570753.7&10.99&0.86&8.93&H&2.35&11.11&Disk&tidal feature - possible SW companion \\
19&SWIRE3\_J104948.86$+$573458.2&10.89&\nodata &\nodata&C/S/L:&2.62&11.24&Disk&edge-on with bright nucleus\\
11&SWIRE3\_J103515.83$+$573337.4&$10.87\pm0.12$ &\nodata &\nodata &\nodata&$>2.96$&11.04&Disk&bright star to N\\
21&SWIRE3\_J105041.96$+$570706.7&10.76&2.08&\nodata &H&2.53&11.19&Disk&edge-on\\
29&SWIRE3\_J105207.16$+$570745.5&10.73&0.11&9.08&H&2.60&10.72&Disk&\\
37&SWIRE3\_J105318.94$+$572140.7&10.70&0.53&8.95&H&2.28&10.88&Disk&\\
2&SWIRE2\_24\_J103249.4$+$573707&10.67&0.58&8.56&H&2.20&10.74&Disk&\\
34&SWIRE3\_J105301.36$+$570543.1&10.64&0.47&8.64&H&2.38&10.49&Disk&compact with bright nucleus\\
7&SWIRE3\_J103327.90$+$574534.4&10.60&0.71&8.87&H&$>2.74$&11.25&Disk&possible tidal arm\\
14&SWIRE3\_J103539.24$+$574243.9&10.58&0.71&8.89&H&2.58&11.03&Disk&small companion to N\\
27&SWIRE3\_J105150.50$+$573905.7&10.56&\nodata &\nodata&\nodata&2.91&10.95&Disk&edge-on\\
18&SWIRE3\_J103653.50$+$575442.5&10.46&0.52&8.43&H&\nodata&10.58&Disk&compact with possible companion to SW\\
40&SWIRE3\_J105432.31$+$570932.4&10.44&0.74&8.77&H&2.44&10.88&Disk&bright nucleus\\
33&SWIRE3\_J105256.85$+$570825.6&10.40&\nodata &\nodata&L:&2.45&10.80&Disk&edge-on with companion to SE \\
26&SWIRE3\_J105143.75$+$572936.9&10.18&1.87&\nodata&S&2.10&11.09&Disk&edge-on with bright nucleus \\
3&SWIRE3\_J103253.94$+$580633.0&10.17&0.55&8.18&H&\nodata&10.26&Disk&disturbed\\
30&SWIRE3\_J105225.75$+$570153.6&10.16&0.14&8.48&H&2.42&10.04&Disk&blue compact\\
24&SWIRE3\_J105100.42$+$574114.9&10.12&1.41&\nodata&H&$>2.93$&10.56&Disk&face-on \\
38&SWIRE3\_J105320.92$+$571433.2&10.03&\nodata &\nodata&\nodata&$>3.09$&11.47&Disk&face-on, disturbed with bright nucleus \\
20&SWIRE3\_J104956.06$+$571440.4&$9.98^{+0.37}_{-0.46}$&\nodata &\nodata&\nodata&$>3.06$&10.23&Disk&small, edge-on\\
17&SWIRE3\_J103606.48$+$574702.4&9.95&0.55&8.58&H&\nodata&9.70&Disk&small, edge-on\\
22&SWIRE3\_J105052.41$+$573506.9&9.64&0.52&8.43&H&2.92&10.40&Disk&small, edge-on\\
4&SWIRE2\_24\_J103258.0$+$573105&9.55&0.49&9.00&H/C:&$>2.88$&10.47&Disk&edge-on\\
39&SWIRE3\_J105349.60$+$570708.1&8.81&\nodata&\nodata&\nodata&2.92&9.45&Disk&blue compact\\
\enddata

\tablecomments{Sources in the LHNW and LHEX fields listed in order of decreasing $L_{\rm IR}$.  Columns give RA ordered counter number, source ID, infrared luminosity (8-1000$\micron$) in units of $L_{\sun}$, extinction, metallicity, spectral type (H = star-forming, C = star forming + AGN, S = Seyfert, L = LINER), stellar mass estimates in log(M$_\sun$) using the methods described in \citet{ilb10}, `{\it q}-value' of FIR-radio correlation, morphology, and Notes (tidal debris and/or companions).  The spectral types are determined by use of the methods of \citet{kew06}, while the extinction is calculated using the Balmer decrement, and metallicities using the [NII]/[OII] diagnostic of \citet{kew02}.}
\tablenotetext{*}{Sources J103538.8$+$573546 and J103341.28$+$580221.4 suffer from contamination due to partially overlapping foreground galaxies, which prohibits determining accurate masses.   Source J105151.64$+$570935.7 is spectroscopically classified as a Seyfert and is suspected to harbor a QSO, thus the use of  {\it Le Phare}  to determine a mass from the UV-NIR photometry is inappropriate.}
 
\label{morph}
\end{deluxetable}
\clearpage
\end{landscape}

\begin{deluxetable}{rlcrrrrrr}
\tabletypesize{\scriptsize}

\tablecaption{Spectral Type Probabilities  from P1P2 Analysis}
\tablewidth{0pt}

\tablehead{
\colhead{\#}&\colhead{Source}&\colhead{log$(L_{\rm IR})$}&\colhead{P1 Color}&\colhead{P2 Color}&\colhead{SF}&\colhead{Comp.}&\colhead{AGN}&\colhead{No EML}\\
\cline{6-9} \\
\multicolumn{2}{c}{}&($L_{\sun}$)&\multicolumn{2}{c}{}&\multicolumn{4}{c}{Probabilities}
}
\startdata
35&SWIRE3\_J105308.32$+$570645.6&$12.74^{+0.19}_{-0.44}$&0.37&0.06&0.37&0.51&0.06&0.06\\
13&SWIRE2\_24\_J103538.8$+$573546\tablenotemark{*}&$12.66^{+0.05}_{-0.04}$&\tablenotemark{*}&\tablenotemark{*}&\tablenotemark{*}&\tablenotemark{*}&\tablenotemark{*}&\tablenotemark{*}\\
8&SWIRE3\_J103341.28$+$580221.4\tablenotemark{*}&$12.62^{+0.10}_{-0.15}$&\tablenotemark{*}&\tablenotemark{*}&\tablenotemark{*}&\tablenotemark{*}&\tablenotemark{*}&\tablenotemark{*}\\
5&SWIRE2\_24\_J103314.8$+$573110&$12.48^{+0.26}_{-0.27}$&0.35&0.00&0.08&0.19&0.29&0.44\\
1&SWIRE3\_J103237.44$+$580845.9&$12.28\pm0.05$&0.11&0.03&0.54&0.31&0.10&0.04\\
12&SWIRE3\_J103526.79$+$575147.6&12.25&0.05&0.06&0.43&0.53&0.04&0.00\\
16&SWIRE3\_J103603.97$+$574812.5&12.20&-0.51&0.01&0.90&0.03&0.02&0.05\\
28&SWIRE3\_J105151.64$+$570935.7\tablenotemark{*}&12.12&\tablenotemark{*}&\tablenotemark{*}&\tablenotemark{*}&\tablenotemark{*}&\tablenotemark{*}&\tablenotemark{*}\\
10&SWIRE2\_24\_J103358.9$+$572952&12.04&0.33&0.09&0.37&0.51&0.06&0.06\\
15&SWIRE2\_24\_J103557.1$+$572234&$11.92^{+0.04}_{-0.84}$&0.22&0.05&0.37&0.51&0.06&0.06\\
25&SWIRE3\_J105113.41$+$571425.9&11.90&0.49&0.01&0.06&0.33&0.36&0.24\\
23&SWIRE3\_J105056.60$+$571631.2&11.69&-0.10&0.01&0.72&0.25&0.02&0.00\\
9&SWIRE3\_J103358.73$+$574317.1&11.46&0.11&0.07&0.43&0.53&0.04&0.00\\
6&SWIRE3\_J103320.32$+$574913.6&11.37&-0.08&0.03&0.72&0.25&0.02&0.00\\
31&SWIRE3\_J105242.40$+$572444.7&$11.05^{+0.07}_{-0.15}$&-0.38&0.02&0.90&0.03&0.02&0.05\\
36&SWIRE3\_J105314.84$+$574137.6&$11.04^{+0.21}_{-0.15}$&-0.19&0.02&0.72&0.25&0.02&0.00\\
32&SWIRE3\_J105252.76$+$570753.7&10.99&0.11&0.05&0.54&0.31&0.10&0.04\\
19&SWIRE3\_J104948.86$+$573458.2&10.89&0.37&0.07&0.37&0.51&0.06&0.06\\
11&SWIRE3\_J103515.83$+$573337.4&$10.87\pm0.12$ &0.10&-0.02&0.34&0.29&0.24&0.12\\
21&SWIRE3\_J105041.96$+$570706.7&10.76&0.35&0.06&0.37&0.51&0.06&0.06\\
29&SWIRE3\_J105207.16$+$570745.5&10.73&-0.12&0.01&0.72&0.25&0.02&0.00\\
37&SWIRE3\_J105318.94$+$572140.7&10.70&-0.16&0.05&0.72&0.25&0.02&0.00\\
2&SWIRE2\_24\_J103249.4$+$573707&10.67&0.04&0.03&0.54&0.31&0.10&0.04\\
34&SWIRE3\_J105301.36$+$570543.1&10.64&-0.05&0.03&0.54&0.31&0.10&0.04\\
7&SWIRE3\_J103327.90$+$574534.4&10.60&-0.05&0.02&0.54&0.31&0.10&0.04\\
14&SWIRE3\_J103539.24$+$574243.9&10.58&0.16&0.04&0.54&0.31&0.10&0.04\\
27&SWIRE3\_J105150.50$+$573905.7&10.56&0.15&0.04&0.54&0.31&0.10&0.04\\
18&SWIRE3\_J103653.50$+$575442.5&10.46&-0.19&0.03&0.72&0.25&0.02&0.00\\
40&SWIRE3\_J105432.31$+$570932.4&10.44&0.20&0.03&0.54&0.31&0.10&0.04\\
33&SWIRE3\_J105256.85$+$570825.6&10.40&0.32&0.05&0.22&0.47&0.19&0.12\\
26&SWIRE3\_J105143.75$+$572936.9&10.18&0.27&0.05&0.37&0.51&0.06&0.06\\
3&SWIRE3\_J103253.94$+$580633.0&10.17&-0.27&0.02&0.72&0.25&0.02&0.00\\
30&SWIRE3\_J105225.75$+$570153.6&10.16&-0.44&0.01&0.90&0.03&0.02&0.05\\
24&SWIRE3\_J105100.42$+$574114.9&10.12&-0.14&0.01&0.72&0.25&0.02&0.00\\
38&SWIRE3\_J105320.92$+$571433.2&10.03&0.45&0.02&0.22&0.47&0.19&0.12\\
20&SWIRE3\_J104956.06$+$571440.4&$9.98^{+0.37}_{-0.46}$&0.16&0.03&0.54&0.31&0.10&0.04\\
17&SWIRE3\_J103606.48$+$574702.4&9.95&-0.30&0.03&0.72&0.25&0.02&0.00\\
22&SWIRE3\_J105052.41$+$573506.9&9.64&0.07&0.04&0.54&0.31&0.10&0.04\\
4&SWIRE2\_24\_J103258.0$+$573105&9.55&0.29&0.03&0.22&0.47&0.19&0.12\\
39&SWIRE3\_J105349.60$+$570708.1&8.81&0.42&0.03&0.22&0.47&0.19&0.12\\\enddata

\tablecomments{Sources in the LHNW and LHEX fields listed in order of decreasing $L_{\rm IR}$.  Columns give RA ordered counter number, source ID, infrared luminosity (8-1000$\micron$) in units of log($L_{\sun}$), corrected P1 color, corrected P2 color, Star-Forming probability, Composite probability, AGN probability, and No Emission-Line probability calculated by comparing P1 and P2 colors with those of the IRAS sample (see Appendix).}
\tablenotetext{*}{Sources J103538.8$+$573546 and J103341.28$+$580221.4 suffer from contamination due to partially overlapping foreground galaxies, which prohibits determining accurate P1,P2 colors.   Source J105151.64$+$570935.7 is spectroscopically classified as a Seyfert and is suspected to harbor a QSO, thus the use of  P1,P2 color to determine spectral type is inappropriate.}
\label{spectab}
\end{deluxetable}

\end{document}